\documentstyle{article}
\begin{document}

{published: {\sf Nucl.Phys.B, {\bf 621} [PM], (2002), 643-674}}
\vskip 0.3cm
\centerline{ \Large \bf Negative moments} 
\centerline{\Large \bf of characteristic polynomials of random
matrices:}
\centerline{ \large \bf Ingham-Siegel integral}
\centerline { \large \bf as an alternative to Hubbard-Stratonovich
transformation}

\vskip 0.5cm
\centerline{ \large \bf Yan V Fyodorov}

\vskip 0.3cm
\centerline{Department of Mathematical Sciences, Brunel University}
\centerline{Uxbridge, UB8 3PH, United Kingdom}
\vskip 0.3cm

\begin{abstract}
We reconsider the problem of calculating 
 arbitrary negative integer moments of the (regularized)
 characteristic polynomial for $N\times N$
random matrices taken from the Gaussian Unitary Ensemble (GUE).
A very compact and convenient integral representation is found
via the use of a matrix integral close to that considered by Ingham
and Siegel. We find the asymptotic expression
for the discussed moments in the limit of large $N$. The latter is
of interest because of a conjectured relation to properties of
the Riemann $\zeta-$ function zeroes.
Our method reveals a striking similarity between the structure of the negative
 and positive integer moments which is usually obscured by the use of the
Hubbard-Stratonovich transformation. This sheds a new light on 
"bosonic" versus "fermionic" replica trick and has some implications for
the supersymmetry method.  We briefly discuss the case
of the chiral GUE model from that perspective. 
\end{abstract}

\section{Introduction}

Recently there was an outburst of research activity related to investigating
the moments and correlation functions of characteristic polynomials
$Z_N(\mu)=\det{\left(\mu {\bf 1}_N-\hat{H}\right)}$ for random 
$N\times N$ matrices $H$ of
various types\cite{zeta1,zeta2,zeta3,BH,Gan,GK,U1,U2,F,MN,KM,LY,AS,FK99,AGG,TS,Ket}.

There are several, not completely independent, sources of motivation
behind studying characteristic polynomials. 
First is the intriguing conjecture relating
limiting distribution of the non-trivial zeroes
$s_k=\frac{1}{2}+it_k$ of the Riemann zeta
function $\zeta(s)$, on the scale of their
mean spacing, to that of (unimodular) eigenvalues of large random
unitary matrices . This implies that locally-determined statistical
properties of $\zeta(s)$, high up the critical line $\mbox{Re}s=1/2$,
might be modelled by the corresponding properties of $Z(\mu)$,
averaged over the so-called Circular Unitary
Ensembles (CUE), i.e. with respect to the normalized Haar measure of 
the group $U(N)$ of $N\times N$ unitary matrices. 
Such a line of thought and
underlying evidences in favour of the conjecture are explained in 
detail in the papers by Keating and collaborators, see
 \cite{zeta1,zeta2,zeta3}.  
In particular, in \cite{zeta1} the authors 
managed to evaluate arbitary moments of
$\left|Z_N\left(\mu=e^{i\theta}\right)\right|$ explicitly:
\begin{equation}\label{zetaf}
{\cal M}_{N,2n}=
\left\langle\left|Z_N\left(\mu=e^{i\theta}\right)
\right|^{2n}\right\rangle=
\prod_{j=1}^N\frac{\Gamma(j)\Gamma(j+2n)}{(\Gamma(j+n))^2}\quad,\quad
\lim_{N\to \infty}\frac{{\cal M}_{N,2n}}{N^{n^2}}=\prod_{j=0}^{n-1}
\frac{j!}{(j+n)!}
\end{equation}
where $\Gamma(z)$ is the Euler gamma-function.
The moments as above were
derived for $\mbox{Re}\,\,{n}>-1/2$ but can be analytically continued
 to the whole complex $n-$plane. The limiting value is presented for
the integer positive $n$. 
This should be compared with the conjecture\cite{zeta1,gon1}
\begin{equation}\label{go1}
\frac{1}{T}\int_0^T dt \left|\zeta(1/2+it)\right|^{2n}\sim a_n\prod_{j=0}^{n-1}
\frac{j!}{(j+n)!} (\log{\frac{T}{2\pi}})^{n^2}
\end{equation}
for the values of the positive integer 
moments of the Riemann $\zeta-$function as $T\to \infty$.
Here $a_n$ is the number specific for
$\zeta$-function\cite{zeta1}, but the rest shows universal 
features common to both the
random matrix calculations and $\zeta-$ function. The parameter 
$\frac{1}{2\pi}\log {(\frac{T}{2\pi})}$ in the above equation 
plays the role of the inverse spacing between
the $\zeta-$ function zeroes at a height $T$ and should be identified
with $N/2\pi$ of the unitary random matrix calculations\cite{zeta1}. 

It is important to have in mind a high degree of {\it universality}
of the obtained results, as discussed in the work by Brezin and Hikami\cite{BH}. By universality one usually means insensitivity
of the spectral characteristics to details of distributions of matrix entries.
In particular, the limiting value Eq.(1) of the positive integer moments of the characteristic polynomials for {\it unitary} matrices is shared, after an appropriate normalisation,
by a broad class of {\it Hermitian} random matrices, whose most prominent representative is the Gaussian Unitary Ensemble (GUE).

Other quantities like the
distribution of the logarithm of the characteristic polynomial, its
derivative, etc. enjoyed thorough
investigations as well\cite{zeta2,zeta3,BH,Gan}. The results were also
extended to the ensemble of unitary symmetric matrices (COE- Circular
Orthogonal Ensemble), which are related to statistics of zeroes of 
the so-called $L-$ functions\cite{L}. An updated summary of open questions on 
relations between the properties of Riemann zeta function and random matrices
can be found at web-page of the American Institute of Mathematics\cite{web}.

What concerns negative moments of the 
characteristic polynomials, an additional interest in calculating 
them arose because of
 a conjectured behaviour of the negative moments
of the (regularized) Riemann zeta function:   
\begin{equation}\label{go2}
\frac{1}{T}\int_0^T dt \left|\zeta\left(1/2+\frac{\delta}{\log{T}}+ 
it\right)\right|^{-2n}\sim  
\left(\frac{\log{\frac{T}{2\pi}}}{\delta}\right)^{n^2}\quad,\quad T\to\infty
\end{equation}
 put forward in \cite{neg} for $1\le \delta \ll \log{T}$.

 The formula Eq.(\ref{zetaf}) shows
divergency at negative integers $n$ and thus provides one with no
explicit answer. Such a divergency is a natural consequence of
necessity to regularize characteristic polynomials by adding a small
imaginary part to the spectral parameter $\mu$ to avoid singularities
due to eigenvalues. When such an imaginary part is comparable with
the separation between the neighbouring eigenvalues one again might expect
universality of the corresponding expressions. 

The Section 8 of the work by Brezin and Hikami \cite{BH} 
discusses a possible way of calculating 
the negative moments of the characteristic polynomials themselves. 
However, in contrast to the moments of the absolute values
those are not divergent and, when taken alone, are insufficient
for the sake of comparison with Eq.(\ref{go2}).

Original goal of the present paper was to reconsider the problem of
calculating both the negative integer moments of the characteristic
polynomials and those of their absolute value.
We succeeded in the analysis of correlation functions of
the (regularized) characteristic polynomials in the limit of 
large $N$ and obtained:
\begin{equation}\label{result}
\lim_{N\to
\infty}\frac{\left\langle\left[Z_N(\mu_1)Z_N(\mu_2^*)\right]^{-n}
\right\rangle}
{\left\langle[Z_N(\mu_1)]^{-n}\right\rangle
\left\langle [Z_N(\mu_2^*)]^{-n}\right\rangle}
=
\left[\frac{2\pi\rho(\mu)}{-i(\mu_1-\mu_2^*)} \right]^{n^2}
\end{equation}
where the regularizations $\mbox{Im}\mu_1>0, \mbox{Im}\mu_2>0$ 
as well as
the spectral difference $\omega=\mbox{Re}{(\mu_1-\mu_2)}$ 
were considered to be of the order of mean eigenvalue spacing
$\Delta_{\mu}=[N\rho(\mu))^{-1}]$, with $\rho(\mu)$ being the mean
eigenvalue density at $\mu=\frac{1}{2}\mbox{Re}(\mu_1+\mu_2)$.

Such an expression complements that given in Eq.(\ref{zetaf}) and is
expected to be universal and applicable to the Riemann zeta-function. 
 Indeed, taking into account the 
nonuniformity of the spectral density for GUE the 
correspondense between the parameters
should be as follows: 
$\Delta_{\mu}^{-1}=N\rho(\mu)\sim \frac{1}{2\pi}\log {(T/2\pi)}$.  
We see that the random matrix result Eq.(\ref{result}) 
and the conjectured Riemann $\zeta$-function behaviour Eq.(\ref{go2}) 
agree in the overall parametric dependence. 

Another motivation for such a calculation comes from some questions that arose
in applications of the random matrix theory to chaotic and disordered 
quantum systems which we shortly discuss below. 

As is well known, eigenvalues of large random matrices\cite{Mbook,GMGW98} 
played a prominent role in the development of the field of quantum
chaos, see e.g.\cite{Haake}, and in revealing its connections to 
mesocopic systems \cite{meso} as well as to some aspects of Quantum
Chromodynamics\cite{QCD}.   
Results on moments and correlation functions of the characteristic polynomials of large random matrices related to 
various aspects of quantum chaotic
systems can be found in \cite{AS,U1,U2,FK99,F}, see also
\cite{AGG,Ket}  for related studies. Characteristic polynomials 
for {\it chiral} random matrix ensembles  are used as a model
partition function for the phenomenon of chiral symmetry breaking
 and as such enjoyed thorough considerations, see
\cite{QCD} and references therein. 
 
Intimately connected with the field of quantum chaos is the domain
of mesoscopic disordered systems. The paradigmatic
example is a single non-relativistic quantum particle moving at zero
temperature in a static random potential. The system Hamiltonian is, in
essence, equivalent to a matrix with random entries. Moreover, in
some limiting case such a matrix belongs to the "domain of
universality" of the classical random matrix theory. 
 This fact is of paramount importance and follows from
the seminal Efetov's work, see the book \cite{Efetov}, 
where the notion of the supermatrix (graded) non-linear $\sigma-$model was introduced for the
first time. The latter  tool alternative to other
techniques in the theory of random matrices expresses expectation
values of the (products of) resolvents of random operators in terms
of integrals over graded matrices containing both commuting and 
anticommuting entries. The method proved to be
 capable of dealing with quantities less accessible by other
methods and turned out to be indispensable in establishing links
between the theory of random matrices and quantum chaotic/mesoscopic
systems, see\cite{Efetov,sasha} and references therein.

An alternative technique which enjoyed many applications in
theoretical physics of disordered systems of interacting particles
 is the (in)famous "replica
trick". Suppose one likes to calculate the ensemble average 
$\langle\log{Z}\rangle$ of a logarithm of some quantity $Z$. The replica trick
exploits the
relation: $\log{Z}=\lim_{n\to 0}\frac{1}{n} (Z^n-1)$ and attempts
 to extract the  averaged logarithm from the behaviour of the moments
$\langle Z^n\rangle$, with $n$ being either positive or 
negative integer.
It is clear that in general the limiting procedure suffers from 
non-uniqueness of the analytical continuation and 
"mathematicians will throw up their hands
in horror or despair, while physicists are much intrigued"\cite{zirn}.   
Random matrices provide an important testing
ground for the replica calculations, with the absolute value 
$|Z_N(\mu)|$ of the characteristic polynomial playing
the role of $Z$. The advantage here is that one has a better control
on results obtained by the ill-defined recipe 
comparing them against those known from independent calculations. 

In particular, the early paper by Verbaarschot and Zirnbauer \cite{VZ} 
devoted to the relation between the replica and supermatrix methods revealed
inherent problems in the former absent in the latter. 
They found that the natural analytic continuation $n\to 0$ gave two
different answers for the "fermionic" (positive moments) and "bosonic" (negative moments") versions of the replica trick, neither of
them coinciding with the known result. In contrast, the latter is correctly reproduced within the supermatrix approach. 

Very recently the verdict of inadequacy of the fermionic replicas
was challenged by Kamenev and Mezard\cite{KM} and further elaborated by
Yurkevich and Lerner\cite{LY}. In particular, Kamenev and Mezard discovered a
convenient integral representation for the integer positive moments 
providing one with a better control on analytical structure of the
expressions. This allowed them to put forward an {\it ansatz} which yielded
in the limit $n\to 0$ the 
correct exact (nonperturbative) result for the GUE matrices, and
the correct asymptotic results for other symmetry classes .
A critical analysis by 
Zirnbauer \cite{zirn}  demonstrated in a coherent manner that
the proposed {\it ansatz} was in no way a well-behaved analytical
continuation. Even so, such a critique did not devaluate the recipe itself but
rather restricted its domain of applicability to perturbative calculations
and called for further investigations.
And indeed, the amended fermionic replica trick immediately 
found applications in the theory of disordered 
 electronic systems with interactions\cite{KM1,YL1} when it was among 
very few tools actually available. Let us also mention a recent
development in the framework of the
Calogero- Sutherland model inspired by closely related ideas\cite{GK}.

The discussed new insights in the nature of the fermionic replica
left, however, unclear if one could come forward with a meaningful amendment
for their bosonic counterpart within the context of nonlinear $\sigma$-model
ideas (see, however, \cite{Kan} for  the replica limit in the context of
orthogonal polynomials). 

An additional motivation for the present paper was to try to bridge the gap
between the cases of the positive and negative $n$. 
Our attempt succeded in discovering an integral representation 
for the {\it negative} integer moments which is strikingly close to that 
obtained by Kamenev and Mezard \cite{KM} for the {\it positive} ones.

Technically, analyticity properties inherent in the negative moments of the
absolute value of characteristic polynomials is known to result in the 
non-compact ("hyperbolic") nature of the integration
manifold for the bosonic nonlinear $\sigma-$model discovered by
Sch\"{a}fer and Wegner\cite{SW}.
In standard considerations such a manifold enters via 
the so-called Hubbard-Stratonovich transformation (see the Appendix D
for more details). 
It came as quite a surprise to the present author that 
the Hubbard-Stratonovich transformation turned out to be not only unnecessary, but
played, in fact, a misleading role hiding the simple structure of the
negative moments. To reveal that structure one should introduce an
alternative route via use of the matrix integral close to one
considered by Ingham\cite{Ingham} and Siegel\cite{Siegel} many years ago.

As to the replica limit, the fact of close similarity between our integral 
representation and those in \cite{KM,LY} 
makes it apparent that very the same KMYL recipe "works"
 for the bosonic version in the same way as 
for its fermionic counterpart.
This should not be considered as contradicting the
Zirnbauer's argumentation since {\it both} versions of the replica trick are
somewhat deficient, in the strict mathematical sense. 
 The result obtained just indicates that accepting one of 
them we have little reasons for discarding the other. 

Clearly, our way of dealing with negative moments suggests
 certain revision of the supermatrix method whose underlying 
technical idea is a simultaneous uniform treatment of
both types of the moments (positive and negative). In fact, we show that
after the disorder average is performed treating "fermionic" and "bosonic" sectors differently can be of some advantage. 

The structure of the paper is as follows. In the section II we expose our method
on the simplest example of negative integer moments of the
characteristic polynomials and analyse the obtained expressions in the limit
$N\to \infty$. Then in the section III we proceed through the calculation for
the negative moments of the absolute value of the polynomial (in fact, a
correlation function). In the section IV we comment on the replica trick and
illustrate our statements by addressing briefly the case considered in
\cite{Ver} - the chiral GUE model - from that perspective. 
Finally, in the section V we present the simplest nontrivial  example of  extention of our method 
to the general type of the correlation (generating) function
containing simultaneously both positive and negative 
moments of  the characteristic polynomials of GUE/chiral GUE matrices.  
The open questions are summarized in the Conclusion.
Technical details are presented in the appendices.

\section{Negative Moments of the Characteristic Polynomial} 
Let $\hat{H}$ be $N\times N$ random Hermitian matrix with
characterized by the standard (GUE) joint probability density:

\begin{equation} \label{GUE}
{\cal P}(\hat{H})=C_N\exp{-\frac{N}{2}\mbox{Tr}\hat{H}^2},\quad 
C_N=(2\pi)^{-\frac{N(N+1)}{2}}N^{N^2/2}
\end{equation}
with respect to the measure 
$d\hat{H}=\prod_{i=1}^N dH_{ii}\prod_{i<j}dH_{ij}d{H^*}_{ij}$. 
Here we use $^*$ to denote complex
conjugation and denote: $dzdz^*\equiv 2d\mbox{Re}z d\mbox{Im} z$.

Regularizing the characteristic polynomial 
$Z_N(\mu)=\det\left(\mu {\bf 1}_N-\hat{H}\right)$
by considering the spectral parameter $\mu$ such that $\mbox{Im}\mu>0$
 one represents negative integer powers of the determinant as
 the Gaussian integral:
\begin{equation} \label{Gau}
[Z_N(\mu)^{-n}]
=\frac{1}{(4\pi i)^{nN}}\int\prod_{k=1}^n
d^2{\bf S}_k
\exp\left\{\frac{i}{2}\mu\sum_{k=1}^n {\bf S}^{\dagger}_k
{\bf S}_k-\frac{i}{2}\sum_{k=1}^n {\bf S}^{\dagger}_k\hat{H}
{\bf S}_k\right\}
\end{equation}
where for $k=1,2,...,n$ we introduced complex $N-$dimensional
vectors
${\bf S}_k=(s_{k,1},...,s_{k,N})^T$ so that 
$d^2{\bf S}_k=\prod_{i=1}^N ds_{k,i}ds^*_{k,i}$
and $T,\dagger$ stand for the transposition and Hermitian
conjugation, respectively.

Denoting by $\left\langle ...\right\rangle$ the expectation value
with respect to the distribution Eq.(\ref{GUE})
we are interested in calculating the negative integer
 moments of the two types:
\begin{equation}\label{mom}
{\cal K}_{N,n}^{(1)}(\mu_1)=\left\langle [Z_N(\mu_1)]^{-n}\right\rangle\quad
\end{equation}
as well as
\begin{equation}
{\cal K}_{N,n}^{(2)}(\mu_1,\mu_2)=\left\langle
\left[Z_N(\mu_1)Z_N(\mu_2^*)\right]^{-n}\right\rangle
\end{equation}
assuming the regularization $\mbox{Im}(\mu_1)=\mbox{Im}(\mu_2)>0$.
In particular, when $\mbox{Re}\mu_1=\mbox{Re}\mu_2$, the latter
quantity amounts to the negative moment of the absolute value
of the characteristic polynomial.

Let us start our consideration with the simplest of the two. 
Performing the ensemble averaging in the standard way one finds
 for the moments of the first type:
\begin{equation}\label{avmom1}
{\cal K}_{N,n}^{(1)}(\mu_1)=
\frac{1}{(4\pi i)^{nN}}\int\prod_{k=1}^n
d^2{\bf S}_k
\exp\left\{\frac{i}{2}\mu_1\sum_{k=1}^n {\bf S}^{\dagger}_k
{\bf S}_k-\frac{1}{8N}\sum_{k,l=1}^n 
\left({\bf S}^{\dagger}_k{\bf S}_l\right)
\left({\bf S}^{\dagger}_l{\bf S}_k\right)\right\}
\end{equation} 

Further introducing a $n\times n$ Hermitian matrix $\hat{Q}$
with the matrix elements $\hat{Q}_{kl}={\bf S}^{\dagger}_k{\bf S}_l$
 the integrand is conveniently rewritten as:
\[
\exp\left\{\frac{i}{2}\mu_1 \mbox{Tr} \hat{Q}-\frac{1}{8N}
\mbox{Tr}\hat{Q}^2\right\}
\]

The standard trick suggested to deal with the 
apparent problem of the non-Gaussian integral above
 is to employ the famous Hubbard-Stratonovich
transformation amounting to:
\begin{equation}\label{HS}
\exp\left\{-\frac{1}{8N}\mbox{Tr}\hat{Q}^2\right\}=\int d\tilde{Q}
\exp\left\{-\frac{N}{2}\mbox{Tr}\tilde{Q}^2-
\frac{i}{2}\mbox{Tr}\tilde{Q}\hat{Q}
\right\}
\end{equation}
thus trading the integration over  $n\times n$ Hermitian matrices 
$\tilde{Q}$ for a possibility to perform the Gaussian integration
over the vectors ${\bf S}_k$. Then the resulting matrix integral is amenable
to the saddle-point treatment in the limit $N\to
\infty$. 

However one may notice a possibility of an alternative 
route. Its starting point is similar to the method employed in \cite{Heu2,LSSS}
where it was suggested to rewrite the integral Eq.(\ref{avmom1}) introducing
the matrix $\delta-$distribution as the product of
$\delta$-distributions of all relevant matrix elements.  
Then, obviously,
\begin{equation}
{\cal K}_{N,n}^{(1)}\propto\int d\hat{Q}e^{-\frac{1}{8N}
\mbox{Tr}\hat{Q}^2} I_n(\hat{Q})
\end{equation}
where 
\begin{equation}\label{I_n}
I_n(\hat{Q})=\int \prod_{k=1}^n
d^2{\bf S}_ke^{\frac{i}{2}\mu_1\sum_{k=1}^n{\bf S}^{\dagger}_k{\bf S}_k}
\prod_{k\le l}
\delta\left(\hat{Q}_{k,l}-{\bf S}^{\dagger}_k{\bf S}_l\right)
\end{equation}
and the $\delta$-distribution for complex variables is understood as the
product of the $\delta$-distributions for their real and imaginary parts.
From now on we do not take care explicitly of multiplicative constants in front
of the integrals. We will show how to restore the constants on a
later stage using the normalisation condition.

To evaluate the last expression we employ the Fourier integral 
representation for each of the delta-functions involved and combine the
Fourier variables into a single $n\times n$ Hermitian matrix
$\hat{F}$. This allows us to proceed as follows:
\begin{eqnarray}\label{Fourier}
I_n(\hat{Q})&\propto& \int \prod_{k=1}^n
d^2{\bf S}_k e^{\frac{i}{2}\mu_1\sum_{k=1}^n{\bf S}^{\dagger}_k{\bf S}_k}
\int d\hat{F}\exp\left\{\frac{i}{2}\mbox{Tr}\left(\hat{F}\hat{Q}\right)
-\frac{i}{2}\sum_{kl}F_{lk}\left({\bf S}^{\dagger}_k{\bf S}_l\right)
\right\}\\
\nonumber
&\propto& \int d\hat{F}
e^{\frac{i}{2}\mbox{Tr}\left(\hat{F}\hat{Q}\right)}
\left[\det{\left(\hat{F}-\mu_1 {\bf 1}_n\right)}\right]^{-N}
\end{eqnarray}

Up to this point our consideration was, in fact, parallel to that employed in 
\cite{Heu2,LSSS}. We however suggest to go one step further by noticing that
the last matrix integral is quite close to the distinguished one considered
originally by Ingham\cite{Ingham} and Siegel\cite{Siegel}:
\footnote{In fact, Ingham and Siegel considered the set of real
symmetric matrices $\hat{F}$ rather than their Hermitian counterparts and 
found the result:
$(\pi)^{\frac{n(n-1)}{4}}\prod_{k=1}^n
\Gamma\left(p+\frac{k+1}{2}\right)\det{M}^{-(p+\frac{n+1}{2})}$.
 However, their method is equally applicable to both cases.}
\begin{equation}\label{IS}
J^{IS}_{p,n}(\hat{Q})=\int_{\hat{F}>0} d\hat{F}
e^{-\mbox{Tr}\left(\hat{F}\hat{Q}\right)}
\left[\det\hat{F}\right]^p
=(2\pi)^{\frac{n(n-1)}{2}}p!(p+1)! ... (p+n-1)!\det{Q}^{-(p+n)}
\end{equation}
where both $\hat{F}$ and $\mbox{Re}\hat{Q}$ are positive definite Hermitian
of the size $n$ and the formula is valid for $p\ge 0$.
 The Ingham-Siegel integral can be viewed as a direct generalisation of 
the Euler gamma-function integral: $\Gamma(p+1)q^{-(p+1)}=\int_{f>0}df f^p e^{-fq}$
to the Hermitian matrix argument and paved a way to 
the theory of special functions of matrix arguments which is nowadays
an active field of research in mathematics and statistics, 
see e.g \cite{matarg}.

It is an easy matter to adopt their method to 
calculating our integral \footnote{We suggest to call such an integral "the Ingham-Siegel integral
of second type".} which is a matrix-argument
generalisation of the formula:$\int_{-\infty}^{\infty}df \frac{e^{ifq}}
{(f-\mu)^N}=\frac{2\pi i}{\Gamma(N)}(iq)^{N-1}e^{if\mu}$ for
$q>0$  and zero otherwise, provided $\mbox{Im}\mu>0$.
Performing the calculation (Appendix A) we find for $N \ge n$:
\begin{equation}\label{theta}
I_{n,N}(\hat{Q}>0)=
\int d\hat{F}e^{\frac{i}{2}\mbox{Tr}\left(\hat{F}\hat{Q}\right)}
\left[\det{\left(\hat{F}-\mu_1 {\bf 1}_n\right)}\right]^{-N}=
C_{N,n}\det{\hat{Q}}^{N-n}e^{\frac{i}{2}\mu_1\mbox{Tr}\hat{Q}}
\end{equation}
with $C_{N,n}=i^{n^2}\frac{(2\pi)^{\frac{n(n+1)}{2}}}{\prod_{N-n+1}^N\Gamma(j)}$ and 
$I_{n,N}(\hat{Q})=0$ whenever at 
least one of the eigenvalues of $Q$ is negative (we recall 
our choice $\mbox{Im} \mu_1>0$).

As a result we arrive 
(after  rescaling the integration variable: $\hat{Q}\to 2N\hat{Q}$)
 to the following integral representation
for the negative integer moments of the characteristic polynomial
in terms of the integral over the matrices $\hat{Q}$:
\begin{equation}\label{k1}
{\cal K}_{N,n}^{(1)}=C_{N,n}^{(1)}\int_{\hat{Q}>0}
 d\hat{Q}e^{-N\left[-i\mu_1\mbox{Tr}{\hat{Q}}+\frac{1}{2}
\mbox{Tr}\hat{Q}^2\right]}\det{\hat{Q}}^{N-n} 
\end{equation} provided $N\ge n$.

 The overall constant $C^{(1)}_{N,n}$ can be restored by
noticing that for $\mbox{Re}\mu_1\to\infty$ the moments tend
asymptotically to $\mu_1^{-nN}$. On the other hand, it is easy to
understand  that such a limit is equivalent to 
discarding the quadratic in $\hat{Q}$ term in the 
exponent of Eq.(\ref{k1}). The resulting integral is precisely the
Ingham-Siegel one, Eq.(\ref{IS}), and comparison yields the required
constant: 
\[
C^{(1)}_{N,n}=(-iN)^{Nn}(2\pi)^{-\frac{n(n-1)}{2}}
\frac{1}{\prod_{j=N-n}^{N-1}j!}
\]

As the last step of the procedure we choose eigenvalues 
$q_1,...,q_n$ and the corresponding
eigenvectors of (positive definite) Hermitian matrix $\hat{Q}$
as new integration variables. This corresponds to the change
of the volume element as: 
$d\hat{Q}=G_n\Delta^2\{\hat{q}\}\prod_{i=1}^n dq_i d\mu(U_n)$ 
where the factor $\Delta^2\{\hat{q}\}=  
\prod_{i<j}(q_i-q_j)^2$ is the squared Vandermonde determinant,
 $G_n=(2\pi)^{\frac{n(n-1)}{2}}
\frac{1}{\prod_{j=1}^{n}j!}$ and
$d\mu(U_n)$ stands for the normalized invariant measure on the unitary group
$U(n)$. The integrand is obviously $U(n)$ invariant and we obtain:
\begin{equation}\label{1}  
{\cal K}_{N,n}^{(1)}(\mu_1)= 
\left\langle\left[\det{(\mu_1 {\bf 1}_N-\hat{H})}\right]^{-n}\right\rangle
=\tilde{C^{(1)}}_{N,n}\int_{q_i>0}
\prod_i\left( dq_i q_i^{-n}\right)\Delta^2\{\hat{q}\}
\exp{-N\sum_{i=1}^n A(q_i)} 
\end{equation}
where 
\begin{equation}\label{action}
\tilde{C}^{(1)}_{N,n}=(-iN)^{Nn}\frac{1}
{\prod_{j=N-n}^{N-1}j!\prod_{j=1}^{n}j!}\quad\mbox{and}\quad
A(q)=\frac{1}{2}q^2-i\mu_1 q-\ln{q}
\end{equation}

The last integral representation is our main result for 
the negative moments of the first type:
${\cal K}_{N,n}^{(1)}(\mu_1)$, valid for arbitary $N>n$. 
One can further play with the formulae for finite $N$ and $n$,
expressing, for example, the negative moments as $n\times n$ determinants:
\begin{eqnarray}
\left\langle\left[\det{(\mu_1 {\bf 1}_N-\hat{H})}\right]^{-n}\right\rangle
&\propto& \det{\left[\Phi_{jk}\right]}\left.\right|^n_{j,k=1}\\
\nonumber \Phi_{jk}&=&\int_0^{\infty}dq q^{N-n}
 \pi^{(1)}_j(q)\pi^{(2)}_k(q)e^{N[i\mu_1 q-\frac{1}{2}q^2]}\quad,
\end{eqnarray}
where $\pi^{(1)}_j(q),\pi^{(2)}_j(q)$ are any monic polynomials
of degree $j$ in the variable $q$, compare with the case of positive moments in \cite{MN}.

In practice, however, we are mostly interested in the limit of large
matrix sizes where one expects the results to show 
universality as was discussed in much detail in the
Introduction. To extract the leading asymptotics as 
$N\to \infty$ when keeping moment
order $n$ fixed one should employ the 
saddle-point method and find
the saddle points of $A(q_i)$. 

Before doing this we observe that the structure of
the derived expressions show striking
similarity to those obtained for the {\it positive} moments of
the characteristic polynomials found in \cite{KM}, see also \cite{BH,MN}:
\begin{equation}\label{1f}  
 \left\langle 
\left[\det{(\mu_{1} {\bf 1}_N-\hat{H})}\right]^{n}\right\rangle
={\tilde{C}^{(1)}}_{N,n}e^{\frac{Nn}{2}\mu_1^2}
\int_{-\infty}^{\infty}
\prod_i dq_i \Delta^2\{\hat{q}\}
\exp{-N\sum_{i=1}^n A(q_i)} 
\end{equation}
where 
\[
\tilde{C}^{(1)}_{N,n}=(-i)^{Nn}N^{n^2/2}\frac{1}{(2\pi)^{n/2}}
\frac{1}{\prod_{j=1}^{n}j!}
\]
and the expression for $A(q)$ is the same as in Eq.(\ref{action}).

The only essential difference between the two representations
 (apart from that in the multiplicative constants and a slight  
change of the power of the determinant: $N-n$ rather than just $N$, 
which is anyway irrelevant for large $N$) is the range of integration. 
For the positive moments one integrates
over the whole real
axis $-\infty<q_i<\infty$ whereas it is over the positive 
semiaxis  $0<q_i<\infty$ for the negative moments.

Thus, we need to consider
the saddle points of $A_{\pm}(q)$.
It is convenient for further reference to define 
$\mu_1=\mu+\frac{\omega}{2}+i\delta$, with $\mu,\omega,\delta$ -real,
 and consider $N\omega,N\delta$
to be fixed when $N\to \infty$. 
Then one can replace $\mu_1$ with $\mu$ in the saddle-point
calculations. The saddle points are obviously given by equations:
\begin{equation}\label{s.p.e}
q_{i}-i\mu-\frac{1}{q_{i}}=0\quad 
\end{equation}
where $i=1,2,...,n$. Each of these equations has two solutions:
\begin{equation}\label{s.p.s.}
q^{\pm}=\frac{i\mu\pm\sqrt{4-\mu^2}}{2}
\end{equation}
We would like to choose the spectral parameter $\mu$
to satisfy $|\mu|<2$ in accordance with the idea of considering
 the bulk of the spectrum for GUE matrices of large size. 
Then only for $q^{+}$ the real parts are positive
and the corresponding saddle points 
contribute to the integral over the positive
semiaxis: $q>0$. 
Consequently, among $2^{n}$ possible sets of 
saddle points $\left(q^{\pm}_{1},...,q^{\pm}_{n}\right)$
only the choice 
\begin{equation}\label{q+}
\hat{q}^+=\mbox{diag}(q_1^+,...,q_1^+)
\end{equation}
should be considered as relevant. This feature
constitutes a considerable difference from the case of positive
moments 
where {\it all} $2^n$ saddle-points yield, in principle, non-trivial
contributions, albeit of different order of magnitude in powers of the small
parameter $N^{-1}$. For example, for $n=2K$ the leading order contribution in the later 
case comes from the choice of half of saddle-points  to be
$q^{+}$, the rest being $q^{-}$, with the combinatorial factor 
$\left(\begin{array}{c}2K\\K\end{array}\right)$ counting the 
number of such sets\cite{BH}.

Presence of the Vandermonde determinants makes the integrand  
vanish at the saddle-point sets of the exponent and thus care should
be taken when calculating the saddle point contribution to the
integral. This part of the procedure uses explicitly the so-called Selberg
integral:
\begin{equation}\label{Sel}
{\cal Z}_n({t})=\int_{-\infty}^{\infty}\prod_{k=1}^n d{\xi_k}
\prod_{k_1<k_2}(\xi_{k_1}-\xi_{k_2})^2 
e^{-\frac{t}{2}\sum_{k=1}^n\xi^2_k}=(2\pi)^{n/2}t^{-n^2/2}\prod_{j=1}^n j!
\end{equation}
for $t>0$,  see the paper by Kamenev and Mezard 
\cite{KM} for more details. General points of their analysis are 
applicable for our case without any modification. 

Expanding around the relevant saddle-points: $q_k=q^{+}+\xi_k$ and
performing the required calculations we find in a
 straightforward way the asymptotic expressions for the negative moments:
\begin{eqnarray}\label{k1+}
{\cal K}_{N,n}^{(1)}(\mu_1)&=&(-i)^{Nn}N^{Nn-\frac{n^2}{2}}
\frac{(2\pi)^{n/2}}{\prod_{j=N-n}^{N-1}j!}
\left[\frac{i\mu+\sqrt{4-\mu^2}}{2}\right]^{Nn+\frac{n^2}{2}}
(4-\mu^2)^{-\frac{n^2}{4}}\\ \nonumber
&\times&\exp\left\{\frac{i\omega N n}{4}(i\mu+\sqrt{4-\mu^2})-\frac{Nn}{2}
\left(1+\frac{\mu^2-i\mu\sqrt{4-\mu^2}}{2}\right)\right\} 
\end{eqnarray}

The formula for ${\cal K}_{N,n}^{(1)}(\mu^*_2)$ where 
$\mu_2^*=\mu-\frac{\omega}{2}-i\delta$ can be obtained
from the above expression by taking its complex conjugate 
and changing $\omega\to -\omega$. Taking the product of the two
expressions we finally find:
\begin{eqnarray}\label{fin1}
{\cal K}_{N,n}^{(1)}(\mu_1){\cal K}_{N,n}^{(1)}(\mu^*_2)
=N^{2Nn-n^2}\frac{(2\pi)^{n}}{\left[\prod_{j=N-n}^{N-1}j!\right]^2}
\left[2\pi\rho(\mu)\right]^{-n^2}\exp\left\{Nn\left[i\pi\rho(\mu)\omega-
\left(1+\frac{\mu^2}{2}\right)\right]\right\} 
\end{eqnarray}
where we used the known expression for the (semicircular) mean
density of GUE eigenvalues: $\rho(\mu)=\frac{1}{2\pi}\sqrt{4-\mu^2}$.

This completes the calculation of the denominator in the formula
Eq.(\ref{result}). To find the corresponding numerator we proceed to 
derivation of the analogous
expressions for the moments of the second type.

\section{Correlation functions for the negative moments of 
 the characteristic polynomials.}

To this end, we consider the product of the expression
Eq.(\ref{Gau}) with its complex conjugate at a different value of the
spectral parameter and average it over the
GUE probability density. From now on we use the index 
$\sigma=1,2$ to label the 
N-component vectors ${\bf S}_{\sigma}$ stemming from the first/second
set of the integrals. To write the resulting expression in a compact
form it is again convenient to introduce $2n\times 2n$ Hermitian matrix $\hat{Q}$
with the matrix elements $\hat{Q}^{\sigma_1,\sigma_2}_{kl}=
{\bf S}^{\dagger}_{\sigma_1,k}{\bf S}_{\sigma_2,l}$, with $k$ and $l$
taking the values $1,...,n$.
In terms of such a matrix we have:

\begin{equation}\label{avmom2}
{\cal K}_{N,n}^{(2)}(\mu_1,\mu_2)\propto\int\prod_{k=1}^n
d^2{\bf S}_{1,k}d^2{\bf S}_{2,k}
\exp\left\{\frac{i}{2}\mu_1\sum_{k=1}^n {\bf S}^{\dagger}_{1,k}
{\bf S}_{1,k}-\frac{i}{2}\mu_2^*\sum_{k=1}^n {\bf S}^{\dagger}_{2,k}
{\bf S}_{2,k}-\frac{1}{8N}\mbox{Tr}\left(\hat{Q}\hat{L}\hat{Q}\hat{L}
\right)\right\}
\end{equation} 
where $\hat{L}=\mbox{diag}({\bf 1}_n,-{\bf 1}_n)$.

Again, the standard way is to use a variant of the Hubbard-Stratonovich
transformation Eq.(\ref{HS}) allowing to convert the term 
quadratic in $\hat{Q}$
(quartic in ${\bf S}$) to that linear in $\hat{Q}$ (quadratic in
${\bf S}$) and integrate out the vectors ${\bf S}$. However, presence
of the matrix $\hat{L}$ and the requirement of convergency of the
Gaussian integrals necessitates introducing this time a rather non-trivial 
 domain (the so-called "hyperbolic manifold", \cite{SW}) for the integration 
over $\tilde{Q}$, to make such a "decoupling" well-defined. 
This problem comprehensively discussed e.g. in \cite{VZ,VWZ} 
 makes the whole procedure technically involved. For a good pedagogical introduction see
\cite{Haake}, the outline of the procedure is presented in the Appendix D of 
the present paper. 

For the method suggested in the present paper
 such problem does not arise at all.
The $2n\times 2n$ matrix 
$\hat{Q}$ is a Hermitian positive definite and the whole
procedure at this stage does not require any modification.  
  Employing the Ingham-Siegel integral of second type
yields in this case:    
\begin{equation}\label{k2}
{\cal K}_{N,n}^{(2)}(\mu_1,\mu_2)=C_{N,n}^{(2)}\int_{\hat{Q}>0}
 d\hat{Q}e^{-N\left[-i\mbox{Tr}{\hat{M}\hat{Q}}+\frac{1}{2}
\mbox{Tr}\left(\hat{Q}\hat{L}\hat{Q}\hat{L}\right)\right]}\det{\hat{Q}}^{N-2n}\quad,\hat{M}=
\mbox{diag}(\mu_1{\bf 1}_n,-\mu_2^* {\bf 1}_n)
 \end{equation} provided $N\ge 2n$, with the overall constant  
\[
C^{(2)}_{N,n}=(N)^{2Nn}(2\pi)^{-n(2n-1)}
\frac{1}{\prod_{j=N-2n}^{N-1}j!}
\]

Clearly, such a uniform applicability can be considered as a
technical advantage. Nevertheless hyperbolic structure, in fact,
lurks in the expression above and manifests itself at the next stage.
Namely, equation Eq.(\ref{k2}) differs from its analogue Eq. (\ref{k1})
in one important aspect: it is now of little utility to introduce 
eigenvalues/eigenvectors of $\hat{Q}$ as integration variables.
Rather, it is natural to treat
$\hat{Q}_L=\hat{Q}\hat{L}$ as a new matrix to integrate over.
Such (non-Hermitian!) matrices are just those forming the mentioned
hyperbolic manifold. I find it
 sensible to discuss their properties explicitly in the Appendix B.
They satisfy $\hat{Q}_L^{\dagger}=\hat{L}\hat{Q}_L \hat{L}$, have
all eigenvalues real and can be
diagonalized by a (pseudounitary) similarity transformation:
$\hat{Q}_L=\hat{T}\hat{q}\hat{T}^{-1}$, where
$\hat{q}=\mbox{diag}(\hat{q}_1,\hat{q}_2)$, and
$n\times n$ diagonal matrices $\hat{q}_1,\hat{q}_2$ satisfy: 
$\hat{q}_1>0\,,\,\hat{q}_2<0$.
Pseudounitary matrices $\hat{T}$ satisfy:
$\hat{T}^{\dagger}\hat{L}\hat{T}=\hat{L}$ and form the group $U(n,n)$
("hyperbolic symmetry").  

In fact, a  more convenient way is rather to
block-diagonalize matrices $\hat{Q}_L$ as 
$$
\hat{Q}_L=\hat{T}_0\left(\begin{array}{cc}\hat{P}_1&\\ & \hat{P}_2
\end{array}\right)\hat{T}_0^{-1}\quad,\mbox{where}\quad 
\hat{T}_0\in \frac{U(n,n)}{U(n)\times U(n)}
$$
and $\hat{P}_{1,2}$ are $n\times n$ Hermitian, with eigenvalues
$\hat{q}_{1,2}$, respectively. The integration measure $d\hat{Q}_L$
is given in new variables as \cite{VZ}: $d\hat{Q}_L=d\hat{P}_1d\hat{P}_2 
\prod_{k_1,k_2}\left(q_{1,k_1}-q_{2,k_2}\right)^2 d\mu(T)$ 
where the last factor is the invariant measure on the manifold of
$T-$matrices whose explicit expression is presented for reference
purposes in the Appendix C.

We therefore arrive to the following expression:

\begin{eqnarray}\label{k3}
{\cal K}_{N,n}^{(2)}&\propto&\int_{\hat{P_1}>0}
\int_{\hat{P_1}<0} d\hat{P}_1d\hat{P}_2\,\, I(\hat{M},\hat{P}_1,\hat{P}_2) \\
\nonumber &\times&
\prod_{k_1,k_2}\left(q_{1,k_1}-q_{2,k_2}\right)^2
 \det{\hat{P}_1}^{N-2n}
\det{\left(-\hat{P}_2\right)}^{N-2n}e^{-\frac{N}{2}\mbox{Tr}
\left(\hat{P}_1^2+\hat{P}_2^2\right)} 
\end{eqnarray}
where
\begin{eqnarray}\label{coset}
&&I(\hat{M},\hat{P}_1,\hat{P}_2)=\int d\mu(\hat{T})
\exp\left\{iN\mbox{Tr}\left(\begin{array}{cc}
\mu_1{\bf 1}_n&\\ & \mu_2^*{\bf 1}_n
\end{array}\right)\hat{T}_0
\left(\begin{array}{cc}\hat{P}_1&\\ & \hat{P}_2
\end{array}\right)\hat{T}^{-1}_0\right\}\\
\nonumber
&&\propto \left[-i(\mu_1-\mu_2^*)\right]^{-n^2}
\frac{1}{\prod_{k_1,k_2}\left(q_{1,k_1}-q_{2,k_2}\right)}
e^{iN\mbox{Tr}\left(\mu_1\hat{q}_1+
\mu_2^*\hat{q}_2\right)}
\end{eqnarray}

The calculation of the above integral is presented in the Appendix C.
We see that its value depends only on the eigenvalue matrices
$\hat{q}_{1}$ and $\hat{q}_{2}$. As a final step we change
$\hat{P}_2\to - \hat{P}_2$ and again introduce
those eigenvalues (and corresponding eigenvectors) 
of the Hermitian matrices $\hat{P}_{1}>0$ and $\hat{P}_{2}>0$ as 
the integration variables.
This results in the following expression for the correlation
function of negative moments of the characteristic polynomial:
\begin{eqnarray}\label{2b}  
\nonumber && {\cal K}_{N,n}^{(2)}(\mu_1,\mu_2)=\left\langle 
\left[\det{(\mu_1 {\bf 1}_N-\hat{H})}
\det{(\mu^*_2 {\bf 1}_N-\hat{H})}\right]^{-n}\right\rangle
\\ 
&=&\tilde{C^{(2)}}_{N,n}\left(\frac{1}{-i[\mu_1-\mu_2^*]}\right)^{n^2}
\int_0^{\infty}\prod_i\, dq_{1,i}\,q_{1,i}^{-2n}\,
\Delta^2\{\hat{q_1}\}
\int_0^{\infty}\prod_i\, dq_{2,i}\, q_{2,i}^{-2n}\,
\Delta^2\{\hat{q_1}\}
\\
\nonumber &\times&
\prod_{k_1,k_2}\left(q_{1,k_1}+q_{2,k_2}\right)e^{-N\sum_{i=1}^n 
A_1(q_{1,i})-N\sum_{i=1}^n A_2(q_{2,i})} 
\end{eqnarray}
where 
\begin{eqnarray}\label{actions}
\tilde{C}^{(2)}_{N,n}=
N^{2Nn-n^2}\frac{1}{\prod_{j=N-n}^{N-1}[j!(j-n)!]
\left[\prod_{j=1}^{n}j!\right]^2}
\end{eqnarray}
and
\begin{equation}
A_1(q)=\frac{1}{2}q^2-i\mu_1 q-\ln{q}\quad,\quad 
A_2(q)=\frac{1}{2}q^2+i\mu_2^* q-\ln{q}
\end{equation}
The constant $\tilde{C}^{(2)}_{N,n}$ given above is most easily checked by considering the limit
$\mu_1\gg \mu_2\gg 1$ in both sides of Eq.(\ref{2b}) and using the identity:
\begin{equation}\label{aux}
\int_0^{\infty}\prod_{i=1}^n\, dq_{i}\,q_{i}^{p}\,
\Delta^2\{\hat{q}\} e^{-\beta\sum_{i=1}^nq_i}=\beta^{-n(n+p)}\prod_{j=1}^{n}j!
\prod_{j=p}^{p+n-1}j!
\end{equation}
valid for $p\ge 0$ and $Re{\beta}>0$. Such a formula is an immediate consequence of
Eq.(\ref{IS}) when going to eigenvalues of the
matrix $\hat{F}$ as integration variables and considering: $\hat{Q}=\beta{\bf 1}_n$.

Again we see that the structure of
the derived expressions is strikingly
similar to those obtained for the correlation functions of 
the positive moments of the characteristic polynomials \cite{KM}:

\begin{eqnarray}\label{2f}  
&&
\left\langle \left[\det{(\mu_1 {\bf 1}_N-\hat{H})}
\det{(\mu^*_2 {\bf 1}_N-\hat{H})}\right]^{n}\right\rangle
\\
\nonumber
&\propto&{\tilde{C}^{(2)}}_{N,n}
e^{\frac{Nn}{2}\left[\mu_1^2+(\mu_2^*)^2\right]}
\int_{-\infty}^{\infty}\prod_i dq_{1,i}
\Delta^2\{\hat{q_1}\}
\int_{-\infty}^{\infty}\prod_i dq_{2,i} 
\Delta^2\{\hat{q_1}\}
\\ \nonumber
&\times&
\prod_{k_1,k_2}\left(q_{1,k_1}+q_{2,k_2}\right)e^{-N\sum_{i=1}^n 
A_1(q_{1,i})-N\sum_{i=1}^n A_2(q_{2,i})} 
\end{eqnarray}
where 

\begin{eqnarray}\label{const1}
 \tilde{C}^{(2)}_{N,n}=
\left[\tilde{C}^{(1)}_{N,n}\right]^2
\frac{1}{\left[-i(\mu_1-\mu_2^*)\right]^{n^2}}\,,
\end{eqnarray}
the constant $\tilde{C}^{(1)}_{N,n}$ is defined earlier in Eq.(\ref{1f})
and expressions for $A_1(q),A_2(q)$ are the same as for the negative
moments, Eq.(\ref{actions}).

Now, however, the difference between the domains of integration
has more important consequences. 
Namely, the negative moments of the absolute
value of the characteristic polynomial are truly divergent
for $(\mu_1-\mu_2^*)\to 0$, as represented by the factor
$(\mu_1-\mu_2^*)^{-n^2}$ in the coresponding formula.
 For their positive counterparts such a
singularity is fake and is compensated when performing the
integration along the whole real axis.

Again, we would like to perform the asymptotic analysis for $N\to \infty$.
 As discussed in the Introduction the most 
interesting "local"  universal regime is to 
occur when one keeps the difference 
$\mbox{Re}(\mu_{1}-\mu^*_{2})\equiv \omega$ and the regularisation $\delta$ 
so small as to ensure
$N \mbox{max}\left(\omega,\delta\right)<\infty$ in such a limit, 
whereas $\mu=\mbox{Re} \frac{(\mu_1+\mu_2)}{2}$ 
is kept in the range $|\mu|<2$. To shorten our notations we include the 
regularization $\delta$ into $\omega$, so that
$\mu_{1,2}=\mu\pm\omega/2$. Then we can write:
\[
N\sum_{i=1}^n A_1(q_{1,i})+N\sum_{i=1}^n A_2(q_{2,i})
=\frac{i}{2}N\omega\sum_{i=1}^n (q_{1,i}+q_{2,i})
+N\left[\sum_{i=1}^n A_+(q_{1,i})+\sum_{i=1}^n A_{-}(q_{2,i})\right]\\,\,,
\]
where the functions $A_{\pm}(q)$ are obtained from $ A_{1,2}(q)$
by setting $\mu_1=\mu_2=\mu$. 

The stationary points of $A_{\pm}(q)$ 
which are obviously given by the equations:
\begin{equation}\label{s.p.e1}
q_{1,i}-i\mu-\frac{1}{q_{1,i}}=0\quad \mbox{and}\quad
q_{2,i}+i\mu-\frac{1}{q_{2,i}}=0
\end{equation}
where $i=1,2,...,n$. Each of these two equations has two solutions:
\begin{equation}\label{s.p.s.1}
q_1^{\pm}=\frac{i\mu\pm\sqrt{4-\mu^2}}{2} \quad\mbox{and}\quad
q_2^{\pm}=\frac{-i\mu\pm\sqrt{4-\mu^2}}{2}\quad,
\end{equation}
but only for $q_{1,2}^{+}$ the real parts are positive
and the corresponding saddle points contribute 
to the integral over the positive
semiaxis: $q_{1,i}>0$ or $q_{2,i}>0$. 
Consequently, among $2^{2n}$ possible sets of 
stationary point $\left(q^{\pm}_{1,1},...,q^{\pm}_{1,n},
q^{\pm}_{2,1},...,q^{\pm}_{2,n}\right)$
only the choice 
\begin{equation}\label{q+1}
\hat{q}^+=\mbox{diag}(q_1^+,...,q_1^+,q_2^+,...,q_2^+ )
\end{equation}
should be considered as relevant. 

Taking care of the Vandermonde determinants via the Selberg
integral Eq.(\ref{Sel}) and calculating 
in this way the fluctuations around the chosen saddle points
 we find the asymptotic
expression for the negative moments of the second type:
\begin{eqnarray}\label{num}
&&\left\langle \left[\det{(\mu_1 {\bf 1}_N-\hat{H})}
\det{(\mu^*_2 {\bf 1}_N-\hat{H})}\right]^{-n}\right\rangle\\ \nonumber
&=&(2\pi)^{n}\left(\frac{1}{-i\left[\mu_1-\mu_2^*\right]}\right)^{n^2}
N^{2n(N-n)}\frac{1}{\prod_{j=N-n}^{N-1}j!(j-n)!}
e^{-nN(1+\mu^2/2)+iNn\pi\omega\rho(\mu)}
\end{eqnarray}
which enters the numerator of Eq.(\ref{result}).
Dividing this expression by that presented in Eq.(\ref{fin1}) and taking into account:
\[
\prod_{N-n}^{N-1}\frac{j!}{(j-n)!}\sim N^{n^2}\quad\mbox{as}\,\, N\to \infty
\]
we arrive at the announced formula Eq.(\ref{result}).

\section{Replica limit. Chiral GUE models}

Let us now briefly consider implications of the derived 
negative moments representations for performing the replica
limit $n\to 0$. We recall the main steps of the scheme 
for the {\it positive} moments ("fermionic replica") as suggested
by Kamenev and Mezard\cite{KM}, see also Yurkevich and Lerner\cite{LY}.

Given the expression Eq.(\ref{2f}), one takes into account two types
of stationary points: the "maximally symmetric" one
$q_{1,i}=q_{2,i}=q^{+}$  as well as all possible sets 
where exactly one of 
$q_{1,i}$ and exactly one of $q_{2,i}$ are taken to be equal to
$q^{-}$, the rest $2n-2$ being equal to $q^{+}$ as before.
It was demonstrated that taking the factors arising from multiplicity of 
the saddle-points and gaussian fluctuations around them
(Selberg integrals) into account only those two possibilities produce leading order
contributions nonvanishing in the replica limit.  

Let us stress clearly the bizarre nature of this prescription, as
compared with the well-defined stationary point procedure for the integer
values $n=1,2,...,$.
First of all, for $n=0$ one takes {\it only one} of two "mostly
symmetric" saddle  points
discarding its partner $q_{1,i}=q_{2,i}=q^{-}$. 
At the same time, for the positive integer moments 
the latter produced exactly
the same contribution  at
vanishing imaginary part $\mbox{Im}\mu\to 0$. 
The step is dictated
by "causality arguments"\cite{KM}, i.e. by necessity to break
analyticity inherent in the positive moments, see \cite{zirn}.

Second, the saddle-point sets containing admixture of two $q^{-}$
contribute now to the same leading order as the "fully symmetric" one, 
whereas for any positive integer $n$ the two contributions were 
different by the factor $1/N$. 
All this is to remind the reader that presently
the replica trick is more a kind of art rather than science (or, rather
 a kind of alchemist's wisdom than regular chemistry). 
For the present author it is however in no way 
an intimidating characteristics but rather a challenge to imagination.

Let us now turn our attention to the negative integer moments as 
described by Eq.(\ref{2b}) and compare them to Eq.(\ref{2f}). 
A little inspection shows that all the factors that make those two
expressions different are immaterial in the replica limit.  
For example, $\lim_{n\to 0}\prod_{N-n}^{N-1}j!
=\lim_{n\to 0}\frac{\prod_{1}^{N-1}j!}{\prod_{1}^{N-n-1}j!}=1$ and
the same is valid for $[\det{\hat{q}}]^n$ and other factors.
 Thus, in
the limit $n\to 0$ the two expressions are indistinguishible 
on the level of saddle point sets
and expansions around them. The only essential deviation which seems to persist is the
difference in the domain of integration, which is half the real axis
for all the negative integer moments. 
The latter feature {\it is} really dictated by
analyticity (or causality),  which, being a
meaningful notion for all negative (but not for the positive!) moments,
dictates only one saddle point to be operative - that with all
$q^{+}$. 

At the same time, there is no obvious reason why other stationary
points should be excluded from a consideration in the replica limit.
All the experience of working with the replicated expressions suggests 
 that saddle-points irrelevant for integer $n$ could be
most relevant for $n=0$, and vice versa. As a distinguished example one can
invoke the famous Sherrington-Kirkpatrick model of spin glasses where
the saddle-points dominating in the replica limit are, in fact, local {\it maxima}
rather than minima of the corresponding functionals. Moreover, formally 
dominant contributions in that case seem to come from 
the boundaries of the integration domains, but are discarded as "unphysical"
 in favour of the mentioned maxima, see e.g.
discussion in p.869 of the reference \cite{spingl}.

We therefore suggest that a
sensible recipe to perform the replica limit for negative moments
of the characteristic polynomials is as
follows: (i) Find an integral representation for the moments 
 with help of the Ingham-Siegel-like integrals (ii) Evaluate the
resulting integral as a sum over the stationary points,
starting with the most symmetric set as dictated by analytical
structure, and adding to it those discovered by
Kamenev and Mezard {\it irrespective} of the constraints on the integration domain.  

To illustrate that such suggestion makes sense beyond the GUE model let us
briefly consider one more example. This is the so-called {\it chiral}
GUE introduced to provide a background for calculating  the universal part 
of the microscopic level density for the QCD Dirac operator, see
\cite{Ver} and references therein. The quantity to be calculated are
{\it negative} moments of the spectral determinant:
\begin{equation}\label{QCD}      
 {\cal I}^{(b)}_{N,n}(m)=A_N \int dJ dJ^{\dagger}
e^{-N\mbox{Tr}\hat{J}^{\dagger}\hat{J}}
\left[\det{\left(\begin{array}{cc}m{\bf 1}_N&i\hat{J}\\ i\hat{J}^{\dagger}&
m{\bf 1}_N \end{array}\right)}\right]^{-n}
\end{equation}
where $m>0$ is a parameter proportional to the quark mass and, 
in the simplest case of zero topological charge, $\hat{J}$ is a 
complex random $N\times N$ gaussian matrix, with $\hat{J}^{\dagger}$
being its conjugate and $A_N$ being the normalisation constant.

We relegate the details of consideration of this interesting and
important model, as well as its close relative - that of
non-Hermitian random matrices -
to a separate publication\cite{unp} and present here only a brief account.  

Application of our method based on the use of the Ingham-Siegel type
integral Eqs. (\ref{IS},\ref{theta}) as an alternative to the conventional 
Hubbard-Stratonovich transformation results in the following
simple formula:
\begin{equation}\label{chir1}
{\cal I}^{b}_{N,n}(m)=C_{n,N}^{ch}\int_{\hat{Q}>0}d\hat{Q}
e^{-mN\mbox{Tr}\hat{Q}}
\left[\det{\left({\bf 1}_n+m\hat{Q}^{-1}\right)}\right]^{-N}
\left[\det{\hat{Q}}\right]^{-n}
\end{equation}
which is exact for arbitrary $N\ge n$. Here $\hat{Q}$ is a positive
definite $n\times n$ Hermitian and the constant is given by: 
$$
C_{n,N}^{ch}=N^{Nn}\frac{1}{(2\pi)^{n(n-1)/2}\prod_{j=N-n}^{N-1}j!}\,\,.
$$
In the thermodynamic chiral limit one considers $m\to 0\,,\,N\to
\infty$ but keeping the product $mN=\frac{x}{2}$ fixed. 
This results in reducing the above expression to: 
\begin{equation}\label{chir2}
{\cal I}^{(b)}_{N,n}(m)=C_{n,N}^{ch}\int_{\hat{Q}>0}d\hat{Q}
e^{-\frac{x}{2}\mbox{Tr}\left(\hat{Q}+\hat{Q}^{-1}\right)}
\left[\det{\hat{Q}}\right]^{-n}
\end{equation}

This should be compared with the corresponding formula for 
the positive moments
\cite{Ver}:
\begin{eqnarray}\label{chir3}
{\cal I}^{(f)}_{N,n}(m)&=&A_N \int dJ dJ^{\dagger}
e^{-N\mbox{Tr}\hat{J}^{\dagger}\hat{J}}
\det{\left(\begin{array}{cc}m\hat{\bf 1}_N&i\hat{J}\\ i\hat{J}^{\dagger}&
m\hat{\bf 1}_N \end{array}\right)}^n\\
\nonumber
&\propto&\int_{U(n)}d\mu(\hat{U})
\exp\{-\frac{x}{2}\mbox{Tr}\left(\hat{U}+\hat{U}^{-1}\right)\}
\end{eqnarray}
where the integration goes over the unitary group $U(n)$.

The correspondence between the two integrals 
is very similar to the GUE case discussed by us earlier in this paper.
Considering $x\to \infty$ as the parameter justifying the saddle-point
approximation one finds the saddle-point sets 
of both integrands coincide: they are given by matrices with 
the eigenvalues $\pm 1$.
Again, in view of the constraint $Q>0$ 
any negative integer moment is dominated by the most symmetric set
with all eigenvalues being equal unity, whereas positive moments are
given by the sum of contributions of many such sets.
 The Kamenev-Mezard-Yurkevich-Lerner limiting procedure which uses
the most symmetric $+1$ configuration as the reference point was
shown to produce sensible results for the integral Eq.(\ref{chir3}) \cite{Ver}.
 Taking into account that the difference between the two integrands
is immaterial in the replica limit, we again arrive at the conclusion
that we can not help but adopt the same scheme for 
proceeding from the negative moments.

%%%%%%%%%%%%%%%%%
\section{Implications for the supersymmetry method}

A curious point which is worth mentioning  is that for the case of {\it positive} moments the
use of the Hubbard-Stratonovich transformation is, in contrast, very natural and
effective. Attempting to use our method for that problem encounters
with the difficulty of dealing with diverging integrals. The latter
hide, in essense, necessity to work with higher
derivatives of the $\delta-$ distributions. 
All this is suggestive of
a certain duality between the two methods: working 
with Grassmann integrations requires the use of the 
Hubbard-Stratonovich transformation but dealing with commuting variables
is much facilitated by avoiding that route. This observation has
 certain implications for the supersymmetry method treating
both types of the variables on equal basis.  From this point  it is of independent
interest to try to treat those two differently, employing the
Ingham-Siegel integrals of second type,  Eq.(\ref{theta}), for the commuting ones.  
Such a treatment for the simplest 
nontrivial  example of  the generating function
containing simultaneously both positive and negative moments 
of  the characteristic polynomials
of GUE matrices is presented
below .  We will see that our method indeed  works and has
certain advantages in comparison with the standard route.  For example, 
the so-called "boundary terms" which 
usually  arise due to a singular nature of the 
transformations along the standard route would not appear in our method.     

We are interested in calculating the following generating function:
\begin{equation}
{\cal C}_{N}(\hat{\mu}_B,\hat{\mu}_F)=
\left\langle
\frac{Z_N(\mu_{1f})Z_N(\mu_{2f})}{Z_N(\mu_{1b})Z_N(\mu^*_{2b})}
\right\rangle_{GUE}
\end{equation}
where $\hat{\mu}_B=\mbox{diag}(\mu_{1b},\mu^*_{2b})\,, \,\,
\hat{\mu}_F=\mbox{diag}(\mu_{1f},\mu_{2f})$ and $\mbox{Im}(\mu_{1b},\mu_{2b})>0$.
That generating function is obviously an analytic one in  $(\mu_{1f},\mu_{2f} )$ plane.
 It  turns out to be technically convenient to change: 
$\mu_{1f}\to -i\mu_{1f}\,\,,\,\,\mu_{2f}\to -i\mu_{2f} $ 
when performing the ensemble averaging, and restore the original generating function
by a simple analytical continuation.

To calculate the average we first use the standard 
"supersymmetrisation" procedure and  represent each of the two 
characteristic polynomials in the denominantor as the Gaussian integrals, Eq.(\ref{Gau}),
whereas those two in the numerator are represented as the Gaussian integrals over the anticommuting
(Grassmannian)  $N-$component vectors ${\bf \chi_1},{\bf \chi_1}^{\dagger}$ and 
$\chi_2,\chi_2^{\dagger}$ , see  e.g. \cite{VZ}. The ensemble average is then easy to perform and
after a straightforward manipulations the generating function can be brought to the 
following form:
\begin{eqnarray}\nonumber
{\cal C}_{N}(\hat{\mu}_b,\hat{\mu}_f)&\propto&
\int d\chi_1 d\chi_1^{\dagger}\int d\chi_2d\chi_2^{\dagger} 
\exp\left\{\frac{1}{2}\left(\mu_{1f}\chi_1^{\dagger}\chi_1+
\mu_{2f}\chi_2^{\dagger}\chi_2\right)
+\frac{1}{8N}\mbox{Tr}\left(\hat{Q}_F^2\right)\right\} 
\\ &\times&
\int d^2{\bf S}_{1} \int d^2{\bf S}_{2}
\exp\left\{\frac{i}{2}\mu_1 {\bf S}^{\dagger}_{1}
{\bf S}_{1}-\frac{i}{2}\mu_2^*{\bf S}^{\dagger}_{2}
{\bf S}_{2}-\frac{1}{8N}\mbox{Tr}\left(\hat{Q}_B\hat{L}\hat{Q}_B\hat{L}
\right)\right\}\\ \nonumber &\times&
\exp\left\{-\frac{1}{4N}\left(\chi_1^{\dagger}\chi_2^{\dagger}\right)\left(
\begin{array}{cc}{\bf S}_{1}\otimes{\bf S}^{\dagger}_{1}-{\bf S}_{2}\otimes
{\bf S}^{\dagger}_{2}& 0\\ 0 &
{\bf S}_{1}\otimes{\bf S}^{\dagger}_{1}-{\bf S}_{2}\otimes{\bf S}^{\dagger}_{2} \end{array}\right)
\left(\begin{array}{c}\chi_1\\ \chi_2\end{array}\right)\right\}
\end{eqnarray} 
where we introduced the $2\times 2$ matrices  
\[
\hat{Q}_F=\left(\begin{array}{cc}\chi_1^{\dagger} \chi_1&
\chi_1^{\dagger} \chi_2\\ \chi_2^{\dagger} \chi_1& \chi_2^{\dagger} \chi_2
\end{array}\right)\quad,\quad 
\hat{Q}_B=\left(\begin{array}{cc}{\bf S}_1^{\dagger} {\bf S}_1&
{\bf S}_1^{\dagger} {\bf S}_2\\ {\bf S}_2^{\dagger} {\bf S}_1& {\bf S}_2^{\dagger} {\bf S}_2
\end{array}\right)
\]
and $\hat{L}=\mbox{diag}(1,-1)$.

Now, following the main  idea outlined in the beginning of the present section we employ 
the Hubbard-Stratonovich identity:
\begin{equation}
\exp{\left[\frac{1}{8N}\mbox{Tr}\hat{Q}_F^2\right]}\propto\int d\tilde{Q}_F
\exp{\left[-\frac{1}{2}\mbox{Tr}\tilde{Q}_F^2-\frac{1}{2\sqrt{N}}
(\chi_1^{\dagger},\chi_2^{\dagger})
\tilde{Q}^T_F\left(\begin{array}{c}\chi_1\\ \chi_2\end{array}\right)\right]}
\end{equation}
where the integration goes over the manifold of $2\times 2$ Hermitian matrices
 $\tilde{Q}_F=\left(\begin{array}{cc}q_{11}&q_{12}\\ q^{*}_{12}& q_{22}
\end{array}\right)$.
Changing the order of integrations and exploiting the above identity one 
performs the (Gaussian) Grassmannian integral explicitly in a simple way and  
brings the expression to the form:
\begin{eqnarray}\label{prom}
&{\cal C}_{N}(\hat{\mu}_b,\hat{\mu}_f)&\propto\\ \nonumber &&
\int d\tilde{Q}_F
e^{-\frac{1}{2}\mbox{Tr}\tilde{Q}_F^2}
\int d^2{\bf S}_{1}d^2{\bf S}_{2}
\exp\left\{\frac{i}{2}\mu_1 {\bf S}^{\dagger}_{1}
{\bf S}_{1}-\frac{i}{2}\mu_2^*{\bf S}^{\dagger}_{2}
{\bf S}_{2}-\frac{1}{8N}\mbox{Tr}\left(\hat{Q}_B\hat{L}\hat{Q}_B\hat{L}
\right)\right\}\\ \nonumber &\times&
\mbox{det}\left(\begin{array}{cc}\left[\mu_{1f}-\frac{1}{\sqrt{N}}q_{11}\right]{\bf 1}_N-
\hat{B} &-\frac{1}{\sqrt{N}}q^*_{12}{\bf 1}_N\\ -\frac{1}{\sqrt{N}}q_{12}{\bf 1}_N& 
\left[\mu_{2f}-\frac{1}{\sqrt{N}}q_{22}\right]{\bf 1}_N-\hat{B}
\end{array}\right)
\end{eqnarray}
where we introduced the $N\times N$ matrix 
$\hat{B}=\frac{1}{2N}
\left[{\bf S}_{1}\otimes{\bf S}^{\dagger}_{1}-{\bf S}_{2}\otimes{\bf S}^{\dagger}_{2}\right]$.

Observing that $\mbox{Tr}\hat{B}^n\propto\mbox{Tr}(Q_B\hat{L})^n$ for any integer $n$,
and introducing $\lambda_{1f},\lambda_{2f}$ as two (real) eigenvalues of the (Hermitian) matrix
$Q_F^{\mu}=\hat{\mu}_F -\frac{1}{\sqrt{N}}\tilde{Q}_F$ one can 
satisfy oneself that the determinant factor in the integrand of Eq.(\ref{prom}) is equal to:
\[
\left(\det{Q_F^{\mu}}\right)^{N-2}\det{\left[\lambda_{1f}{\bf 1}_2  -\frac{1}{2N}\tilde{Q}_B\hat{L}
\right]}\det{\left[\lambda_{2f}{\bf 1}_2  -\frac{1}{2N}\tilde{Q}_B\hat{L}
\right]}
\]
Next step is to evaluate the integral 
over ${\bf  S}_1,{\bf  S}_2$. Since the integrand 
depends on those variables only via the matrix  $\tilde{Q}_B$ 
the calculation can be done
exactly the same way as elsewhere in the paper. 
That  involves the $\delta$-function 
representations and the 
subsequent Ingham-Siegel second type integration, 
see Eqs. (\ref{Fourier},\ref{theta},\ref{k1}). 
The resulting expression is given by:

\begin{eqnarray}\label{prom1}
{\cal C}_{N}(\hat{\mu}_b,\hat{\mu}_f)&\propto& \\ \nonumber 
&&\int d\tilde{Q}_F\left(\det{Q_F^{\mu}}\right)^{N-2}
e^{-\frac{1}{2}\mbox{Tr}\tilde{Q}_F^2}
\int d\hat{Q}_B \det{\hat{Q}_B}^{N-2}
\exp\left\{-\frac{1}{8N}\mbox{Tr}\left(\tilde{Q}_B\hat{L}\right)^2+
\frac{i}{2}\mbox{Tr}\left[\hat{\mu}_B\hat{Q}_B\hat{L}\right]\right\}
\\ \nonumber &\times&
\det{\left[\lambda_{1f}{\bf 1}_2  -\frac{1}{2N}\tilde{Q}_B\hat{L}
\right]}\det{\left[\lambda_{2f}{\bf 1}_2  -\frac{1}{2N}\tilde{Q}_B\hat{L}
\right]}
\end{eqnarray}

Further steps involve: (i) changing: $\tilde{Q}_F\to \sqrt{N}\tilde{Q}_F\, ,\, 
\tilde{Q}_B\to 
{2N}\tilde{Q}_B$ (ii) diagonalizing $\hat{Q}_B\hat{L}=\hat{T}\mbox{diag}(p_1,p_2)\hat{T}^{-1}$,
where $p_1>0, p_2<0$ (iii) integrating over the "hyperbolic" 
manifold of the pseudounitary matrices 
$\hat{T}$ in the same way as it is done in Eqs. (\ref{k3}, \ref{coset}) (iv) shifting the integration  over
 $\tilde{Q}_F$ to that over
$\tilde{Q}^{(\mu)}_F$   and introducing the
 (real) eigenvalues $q_{1},q_{2}$ of the latter matrix 
(and the corresponding unitary matrices of eigenvectors $\hat{U}_F$)
 as new integration variables. (v) performing the integration over the unitary group 
$\hat{U}_F$ by the standard Itzykson-Zuber-Harish-Chandra formula, see Appendix C.
Continuing finally  analytically as $(\mu_{1f},\mu_{2f})\to (i\mu_{1f}, i\mu_{2f})$ we arrive at
\footnote{One should take into account a symmetry of the integrand with respect to the transformation: $q_1\to q_2$ to simplify the expression.}
\begin{eqnarray}\label{prom2}
{\cal C}_{N}(\hat{\mu}_b,\hat{\mu}_f) &\propto& e^{\frac{N}{2}
(\mu_{1f}^2+\mu_{2f}^2)}\frac{1}{(\mu_{1b}-\mu^*_{2b})(\mu_{1f}-\mu_{2f})}
\\ \nonumber &\times&
\int_{-\infty}^{\infty}dq_{1}\int_{-\infty}^{\infty}dq_{2}
\frac{(q_{1}-q_{2})}{(q_{1}q_{2})^2}
\exp\{-N\left[{\cal L}_{1f}(q_{1})+{\cal L}_{2f}(q_{2})\right]\}
\\ \nonumber &\times&
\int_{0}^{\infty}dp_{1}\int_{0}^{\infty}dp_{2}
\frac{(p_{1}+p_{2})}{(p_{1}p_{2})^2}
\exp\{-N\left[{\cal L}_{1b}(p_{1})+{\cal L}_{2b}(p_{2})\right]\}
\end{eqnarray}
where 
\begin{eqnarray}\label{prom3}
{\cal L}_{1f}(q)=\frac{1}{2}q^2-i\mu_{1f}q-\ln{q}\quad,\quad
{\cal L}_{2f}(q)=\frac{1}{2}q^2-i\mu_{2f}q-\ln{q}\\
{\cal L}_{1b}(p)=\frac{1}{2}p^2-i\mu_{1b}p-\ln{p}\quad,\quad
{\cal L}_{1b}(p)=\frac{1}{2}p^2+i\mu^*_{2b}p-\ln{p}
\end{eqnarray}

So far all the formulae were exact for any integer $N\ge 2$.  Let us stress an apparent symmetry
of the resulting equation with respect to $q$ and $p$ variables, apart from the anticipated
difference in the integration domain. 
Consider now the limit $N\to \infty$, where the integrals are dominated by the saddle-points:
\begin{eqnarray}\label{prom4}
q^{\pm}_{1f}=\frac{1}{2}\left[i\mu_{1f}\pm\sqrt{4-\mu_{1f}^2}\right]\quad,
\quad
q^{\pm}_{1f}=\frac{1}{2}\left[i\mu_{1f}\pm\sqrt{4-\mu_{1f}^2}\right]\\ \nonumber
p_{1}=\frac{1}{2}\left[i\mu_{1b}+\sqrt{4-\mu_{1b}^2}\right]\quad,\quad
p_{2}=\frac{1}{2}\left[-i\mu_{2b}+\sqrt{4-\mu_{2b}^2}\right]
\end{eqnarray}
Here we took into account the restrictions $\mbox{Re} p_{1,2}\ge 0$ and also set $ \mbox{Im} \mu_{(1,2)b}=0$.

In fact, in the limit $N\to \infty$ we are interested in the 
"local regime" : $\mu_{(1,2)b}=\mu\pm\frac{\omega_b}{2}\,,\,
\mu_{(1,2)f}=\mu\pm\frac{\omega_f}{2}$ where $|\mu|\le 2$ and $\omega_b\,,\,\omega_f=O(1/N)$. We find it convenient to define:
\[
\mu_{(1,2)f}=2\sin{\phi_{(1,2)f}}\quad,\quad \mu_{(1,2)b}=2\sin{\phi_{(1,2)b}}
\]
Then $\omega_{b,f}= \mu_{1(b,f)}- \mu_{2(b,f)}\sim (\phi_{1(b,f)}-\phi_{2(b,f)})
2\cos{\phi}$, where $2\cos{\phi}=
\sqrt{4-\mu^2}=\pi\rho(\mu)$ is proportional to the mean spectral density. Thus we have:
\[
\phi_{1b}-\phi_{2b}=\frac{\omega_b}{\pi\rho(\mu)}\quad,\quad \phi_{1f}-\phi_{2f}=\frac{\omega_f}{\pi\rho(\mu)}
\]
The saddle-point values, Eqs.(\ref{prom4}) are given by:
\[
p_1=e^{i\phi_{1b}}\quad,\quad p_2=e^{-i\phi_{2b}}\quad,\quad q ^{\pm}_1=e^{\pm i\phi_{1f}}\quad,\quad
q ^{\pm}_2=e^{\pm i\phi_{2f}}
\]
which shows that:
$$
q^{\pm}_1-q ^{\pm}_2=e^{\pm i\phi_{1f}}\left(1-e^{\pm i(\phi_{1f}-\phi_{2f})}\right)=O(1/N)\quad,\quad
q ^{\pm}_1-q ^{\mp}_2=\pm 2\cos{\phi_{(1,2)f}}+O(1/N)\approx\pm\pi\rho(\mu)
$$
The latter relation makes it clear that only two saddle point sets $(q_1^{\pm},q_2^{\pm})$ give dominant contribution whereas the other two:
$(q_1^{\mp},q_2^{\pm})$ are suppressed due to the factor $(q_1-q_2)$ in the integrand of Eq.(\ref{prom2}).

Expanding around the relevant saddle points and picking up the leading order contributions we obtain after a standard set of manipulations:
\begin{eqnarray}\label{prom5}
\nonumber 
{\cal C}_{N\gg 1}(\hat{\mu}_b,\hat{\mu}_f) &\propto&
e^{iN(\phi_{1b}-\phi_{2b})}
\frac{1}{(\mu_{1b}-\mu_{2b})(\mu_{1f}-\mu_{2f})}
\\ \nonumber &\times&
\left\{e^{iN(\phi_{1f}-\phi_{2f})}\left(e^{ i\phi_{1f}}-e^{ i\phi_{1b}} \right)\left(e^{ -i\phi_{2f}}-e^{ -i\phi_{2b}} \right)\right.\\ \nonumber
&&\left.- e^{-iN(\phi_{1f}-\phi_{2f})}\left(e^{ -i\phi_{1f}}-e^{- i\phi_{2b}} \right)\left(e^{ i\phi_{2f}}-e^{ -i\phi_{1b}}\right)\right\}
\\ &\propto&
\frac{1}{\omega_b\omega_f}\left[e^{iN\pi\rho(\mu)(\omega_b+\omega_f)} (\omega_b-\omega_f)^2-
e^{iN\pi\rho(\mu)(\omega_b-\omega_f)} (\omega_b+\omega_f)^2
\right]
\end{eqnarray}
The latter expression coincides with the particular case of the result obtained in \cite{AS} by a different method. 

The method based on the combination of the Ingham-Siegel and Hubbard-Stratonovich
transformations proves also to be a direct way to 
derive the mean density of eigenvalues for the chiral
GUE ensemble discussed in the previous section. 
Let us outline the corresponding calculation
for the (quenched) case of zero topological charge.

The generating function to be calculated is given by (cf. Eq.(\ref{QCD})):
\begin{equation}\label{QCD1}      
 {\cal I}_{N}(m_f,m_b)\propto \int dJ dJ^{\dagger}
e^{-N\mbox{Tr}\hat{J}^{\dagger}\hat{J}}
\frac{\left[\det{\left(\begin{array}{cc}m_f{\bf 1}_N&i\hat{J}\\ i\hat{J}^{\dagger}&
m_f{\bf 1}_N \end{array}\right)}\right]}
{\left[\det{\left(\begin{array}{cc}m_b{\bf 1}_N&i\hat{J}\\ i\hat{J}^{\dagger}&
m_b{\bf 1}_N \end{array}\right)}\right]}
\end{equation}
where $m_f , m_b>0$.

Representing  the determinant in the numerator and that in the denominator as the Gaussian integrals over anticommuting
and commuting variables respectively, we easily perform the ensemble averaging, and then employ the Hubbard-Stratonovich
transformation for the terms quartic in  Grassmannians whereas using the Fourier representation of $\delta$-function
for their commuting counterparts.  This allows to integrate out Grassmann variables exactly and after the Ingham-Siegel integration to arrive to the following expression:
\begin{eqnarray}\label{QCD2}      
 {\cal  I}_{N}(m_f,m_b)&\propto& \int dq dq^* (q^*q)^{N-1}
\exp{-\frac{N}{4}\left[\left\{q^{*}q-\frac{m_f}{2}(q+q^*)+\frac{1}{4}m_f^2\right\}\right]}\\ \nonumber
&\times& \int_{0}^{\infty} dp_1 \int_{0}^{\infty}dp_2
(p_1p_2)^{N-1}\left(q^*q-p_1p_2\right)
\exp{-\frac{N}{4}\left\{p_1p_2+\frac{m_b}{2}(p_1+p_2)\right\}}\end{eqnarray}
Changing now to the polar coordinates $R,\theta$ in the complex $q,q^*$ plane and next to the variable $t=p_1p_2$
one can perform the $\theta$ and $p_1$-integrations explicitly finding:
\begin{equation}\label{QCD3}      
 {\cal  I}_{N}(x_f,x_b)\propto e^{-\frac{x_f^2}{N}}\int_{0}^{\infty} dr \int_{0}^{\infty}dt
(rt)^{N-1}\left(r-t\right)
e^{-N(r+t)}I_0(2x_fr^{1/2})K_0(2x_bt^{1/2})
\end{equation}
where $x_{b,f}=Nm_{b,f}/4$ and $I_0(z) , K_0(z)$ stand for the modified Bessel and Macdonald functions, respectively.
In the most interesting chiral limit we set $N\to \infty$ while keeping $x_{b,f}$ fixed. The saddle-point values are obviously 
given by $r=t=1$ and expanding around them up to the first non-vanishing term  finally yields:
\begin{equation}\label{QCD4}      
 {\cal  I}_{chiral}(x_f,x_b)\propto
\left[x_f I_1(2x_f) K_0(2x_b)+x_b I_0(2x_f) K_1(2x_b)\right]
\end{equation}
in full agreement with known results, see \cite{Ver,AS1}.
  
%%%%%%%%%%%

\section{Conclusions and Perspectives}

In the present paper we suggested a systematic way of evaluating
negative integer moments of the (regularized) characteristic polynomials. 
Using  the standard representation of those moments in terms of the Gaussian
integrals as the starting point we found a route avoiding the use of the ubiquitous
Hubbard-Stratonovich transformation. Instead, we advocated the
exploitation of the matrix integral Eq.(\ref{theta}) similar to that considered long ago by Ingham and Siegel, Eq.(\ref{IS}). The advantage of the
procedure is that the emerging structures are attractively simple 
and, in essense, very close to those derived earlier for the positive
moments. We evaluated the resulting integrals in the limit $N\to
\infty$ by the stationary phase method and extracted 
the leading asymptotics. The limiting value of the correlation
functions for the negative moments is presented in Eq.(\ref{result})
and expressed in terms suggesting universality. In fact, by using
the character expansion method we were able to reproduce the formula (\ref{result}) for unitary random matrices\cite{Str}. We therefore conjecture
that it should be equally applicable, mutatis mutandis, to the behaviour
of the Riemann zeta function in the close vicinity of the critical
line. 
 
Our analysis may raise a few questions which deserve further
discussion. First of all, one may wish to know if it is possible 
to arrive to the same representations via the standard (Hubbard-Stratonovich) method.
The answer is of course affirmative as is demonstrated in the Appendix D on
the simplest non-trivial example. The general case can be treated 
along the same lines. We however insist that such a way is hardly natural for the
present problem and, in fact, obscures the simple structures arising.

This fact has important implications for the "supersymmetry method" 
of treating generating functions involving simultaneously characteristic polynomials both in positive and negative powers. 
We presented a few simple non-trivial examples of such calculations in the text. In fact, we found it natural to treat
"bosonic" and "fermionic" auxilliary integrations differently, which was a serious departure from the standard spirit of "supersymmetry".

One should mention that the correlation functions calculated above can be also calculated in the 
framework of the standard Efetov's supermatrix  approach. However, present method, to our mind, has some 
technical advantages. For GUE calculations, for example, 
the two terms appearing in the final result
(\ref{prom5}) on equal footing have quite a different origin in the Efetov's method. Namely, one of them  appears in the 
form of the so-called "anomalous", or "boundary term". When trying to calculate the correlation functions of the higher orders
the boundary terms turn out to be quite difficult to keep track of,
 making the $\sigma-$model calculation a daunting job.

In fact, the authors of \cite{AS}
announced  the result for the most general expression of the ratios of integer powers of  characteristic polynomials .
They arrived to it by a method generalizing that suggested in the paper by  Guhr\cite{Guhr} and employing 
an analog of the Itzykson-Zuber integral over a supermanifold.  Our method  provides an alternative way of derivation of 
the higher correlation function, the corresponding calculation will be presented in a detailed form elsewhere\cite{unp1}

For the case of chiral GUE ensemble the application of the standard Efetov's method for the quenched model 
requires quite a cumbersome calculation \cite{AS1}, whereas along our route the calculations are rather
short and  elementary.  Moreover,  we found \cite{unp} that the same method proved 
to be applicable for the most interesting "unquenched" situation  when the Efetov's method encountered with 
unsurmountable difficulties.
 
Finally, it is interesting to explore if the Ingham-Siegel integrals
and their natural generalisations could provide a serious alternative
to the Hubbard-Stratonovich transformation in the whole class of
problems in the domain of random matrices and disordered systems.
To this end two aspects are worth mentioning: (i) the Ingham-Siegel
integrals of both types are known for all symmetry classes, see Appendix A and 
(ii) performing the saddle-point calculation directly on the level of
Eq.(\ref{k2}) yields the standard non-linear $\sigma-$model
representation\cite{VZ} for the negative moments.
Further work along these lines is under the way\cite{unp, unp1} but a
general affirmative or negative answer to the questions requires
more efforts.

\section*{Acknowledgements}
The author acknowledges many useful discussions
with Hans-Jurgen Sommers over the years. Those were 
an important input for the present research at early stages of the work.
The author is grateful to G.Akemann, A. Kamenev, D.Savin, H.A. Weidenm\"{u}ller and  J.J.M. Verbaarschot for their encouraging interest in the work and useful comments.

This research was supported by EPSRC grant GR/R13838/01 "Random matrices close to
Unitary or Hermitian".

%%%%%%%%%%%%%%%%%%%
\appendix
{\bf APPENDICES}
\section{ Calculation of the integral Eq.(15)}
Our goal is to calculate the integral
\begin{equation}\label{theta1}
I_{n,N}(\hat{Q_n})=
\int d\hat{F_n}e^{\frac{i}{2}\mbox{Tr}\left(\hat{F_n}\hat{Q_n}\right)}
\left[\det{\left(\hat{F_n}-\mu {\bf 1}_n\right)}\right]^{-N}
\end{equation}
where both $\hat{F}_n$ and $\hat{Q}_n$ are Hermitian $n\times n$ matrices.
First  notice that the integrand is 
invariant with respect to the unitary rotations 
$F\to \hat{U}\hat{F}\hat{U}^{-1}$, hence the result of the integration
can depend only on the eigenvalues of $\hat{Q}$. Then, following
\cite{Ingham,Siegel} one can
take $\hat{Q}$ to be diagonal from the very beginning and 
separate the first eigenvalue from the rest:
\[
\hat{Q}=\mbox{diag}(q_1,q_2,...,q_n)\equiv \mbox{diag}(q_1,\hat{Q}_{n-1})
\]
 Accordingly decompose the
matrix $\hat{F_n}$ as 
\begin{equation}
\hat{F}_n=\left(\begin{array}{cc}f_{11} & {\bf f}^{\dagger}\\
{\bf f}& \hat{F}_{n-1}\end{array}\right)\quad,\quad d\hat{F}_n=df_{11}d{\bf
f}^{\dagger} d{\bf f}d\hat{F}_{n-1}
\end{equation}
where ${\bf f}^{\dagger}=\left(f^*_{21},f^*_{31},....,f^*_{n1}\right)$
is a $n-1$ component complex vector.

Next step is to use the well-known property of the determinants:
\[
\det{\left(\hat{F_n}-\mu {\bf 1}_n\right)}=
\det{\left(\hat{F}_{n-1}-\mu {\bf 1}_{n-1}\right)}
\left(f_{11}-\mu-{\bf f}^{\dagger}
\left[\hat{F}_{n-1}-\mu{\bf 1}_{n-1}\right]^{-1}{\bf f}\right)
\] 
which gives:
\begin{eqnarray}
I_{n,N}(\hat{Q}_n)&=&
\int d\hat{F}_{n-1}e^{\frac{i}{2}\mbox{Tr}\left(\hat{F}_{n-1}\hat{Q}_{n-1}\right)}
\left[\det{\left(\hat{F}_{n-1}-\mu {\bf 1}_{n-1}\right)}\right]^{-N}\\
\nonumber &\times& \int d{\bf f}^{\dagger}d{\bf f}\int_{-\infty}^{\infty}
df_{11}e^{\frac{i}{2}f_{11}q_1}\frac{1}
{\left(f_{11}-\mu-{\bf f}^{\dagger}
\left[\hat{F}_{n-1}-\mu{\bf 1}_{n-1}\right]^{-1}{\bf f}\right)^N}
\end{eqnarray}
The last integral over $f_{11}$ is evaluated by the residue theorem
taking into account $\mbox{Re}\mu>0$, the result of the
 integration being:
\begin{equation}
\frac{2\pi i}{\Gamma(N)}\theta(q_1)\left(\frac{iq_1}{2}\right)^{N-1}
\exp\left\{\frac{i}{2}q_1\left(\mu+{\bf f}^{\dagger}
\left[\hat{F}_{n-1}-\mu{\bf 1}_{n-1}\right]^{-1}{\bf f}\right)\right\}
\end{equation}
where $\theta(x)=1$ for $x>0$ and zero otherwise and we assumed $N\ge
1$. Now the gaussian integration over $d{\bf f}^{\dagger}d{\bf f}$
can be easily performed, yielding the factor:
\[
\left(\frac{-2\pi}{(iq_1/2)}\right)^{n-1}
\mbox{det}\left(\hat{F}_{n-1}-\mu{\bf 1}_{n-1}\right)
\]
so that we arrive at the recurrence relation:
\begin{equation}
I_{n,N}(\hat{Q}_{n})=\frac{(-2\pi)^n(-i)}{\Gamma(N)}\left(\frac{i}{2}q_1\right)^{N-n}
 \theta(q_1)e^{\frac{i}{2}\mu q_1}I_{n-1,N-1}(\hat{Q}_{n-1})
\end{equation}

which immediately produces the desired formula:
\begin{equation}
I_{n,N}(\hat{Q}_{n})=i^{n^2}\frac{(2\pi)^{\frac{n(n+1)}{2}}}
{\prod_{N-n+1}^N\Gamma(j)}
\prod_{j=1}^n\theta(q_j)\mbox{det}\left[\frac{i}{2}\hat{Q}\right]^{N-n}
e^{\frac{i}{2}\mu\mbox{Tr}\hat{Q}}
\end{equation}
assuming $N\ge n$.

In fact, the derivation is straightforwardly repeated for the case of
real symmetric matrices $\hat{F}_n$ and $\hat{Q}_n$.
The recurrence relation in that case is:
\begin{equation}
I^{r.s.}_{n,N}(\hat{Q}_{n})=2\frac{i^n\pi^{\frac{n+1}{2}}}{\Gamma(N)}
\left(\frac{i}{2}q_1\right)^{N-\frac{n+1}{2}}
 \theta(q_1)e^{\frac{i}{2}\mu q_1}I^{r.s.}_{n-1,N-\frac{1}{2}}(\hat{Q}_{n-1})
\end{equation}
which yields the result:
\begin{equation}
I^{r.s.}_{n,N}(\hat{Q}_{n})=2^n\frac{i^{\frac{n(n+1)}{2}}(\pi)^{\frac{n(n+3)}{4}}}
{\prod_{N-\frac{n-1}{2}}^N\Gamma(j)}
\prod_{j=1}^n\theta(q_j)\mbox{det}
\left[\frac{i}{2}\hat{Q}\right]^{N-\frac{n+1}{2}}
e^{\frac{i}{2}\mu\mbox{Tr}\hat{Q}}
\end{equation}
for $N\ge \frac{n+1}{2}$.

%%%%%%%%%%%%%%%%%%%

\section{Properties of the matrices $\hat{Q}_L$}

In this Appendix we consider the manifold of $2n\times 2n$ matrices
$\hat{Q}_L=\hat{Q}\hat{L}$ where $\hat{Q}^{\dagger}=\hat{Q}>0$ and 
$\hat{L}=\mbox{diag}({\bf 1}_n,-{\bf 1}_n)$. In fact, this set of
matrices is closely related to the object known as a regular matrix pensil
\cite{Gant}.

We begin with proving that all eigenvalues of
such non-Hermitain matrices are real and half of them positive, the
rest being negative. 
In doing this we can safely assume that all
eigenvalues are different since matrices with degenerate eigenvalues 
form a manifold of lower dimension and as such will not contribute 
when we integrate over the whole manifold of $\hat{Q}_L$.

 The characteristic polynomial for the eigenvalues $q$ of the matrix
 $\hat{Q}\hat{L}$ can be written as:
\[
\det{\left(q{\bf 1}_{2n}-\hat{Q}\hat{L}\right)}=
\det{\hat{Q}^{1/2}\left(q{\bf 1}_{2n}-\hat{Q}^{1/2}\hat{L}\hat{Q}^{1/2}\right)
\hat{Q}^{-1/2}}=
\det{\left(q{\bf 1}_{2n}-\hat{Q}^{1/2}\hat{L}\hat{Q}^{1/2}\right)}
\]
where we used that $\hat{Q}^{1/2}>0$ is a nonsingular Hermitian matrix.
Then all eigenvalues of $\hat{Q}_L$ coincide with those of the Hermitian
$\hat{Q}^{1/2}\hat{L}\hat{Q}^{1/2}$ and therefore are all real.
Moreover, the number of positive and negative eigenvalues of 
any Hermitian matrix $\hat{H}$ stays invariant under 
transformations $\hat{H}\to \hat{T}^{\dagger}\hat{H}\hat{T}$, where
$\hat{T}$ is an arbitary nonsingular matrix \cite{Gant}. We arrive at the
conclusion that the number of positive and negative eigenvalues of
$\hat{Q}^{1/2}\hat{L}\hat{Q}^{1/2}$ is the same as that for
$\hat{L}$, thus proving the statement.
 
Let $q_j$ be an eigenvalue of $\hat{Q}_L$ and denote the corresponding
(right) eigenvectors as ${\bf e}_j$:
$$
\left(\hat{Q}\hat{L}\right){\bf e}_j=q_j {\bf e}_j\quad,\quad
{\bf e}^{\dagger}_j\left(\hat{L}\hat{Q}\right)=q_j {\bf e}^{\dagger}_j
$$
Multiplying the first of these relations with 
${\bf e}^{\dagger}_k\hat{L}$ from the left and the second relation with 
$\hat{L}{\bf e}_j$ from the right we have:
\[
{\bf e}^{\dagger}_k\left(\hat{L}\hat{Q}\hat{L}\right){\bf e}_j=q_j 
{\bf e}^{\dagger}_k\hat{L}{\bf e}_j=
q_k {\bf e}^{\dagger}_k\hat{L}{\bf e}_j
\]
showing that ${\bf e}^{\dagger}_k\hat{L}{\bf e}_j=\delta_{jk}
q_j {\bf e}^{\dagger}_j\hat{L}{\bf e}_j$ 

Now, $\hat{L}$ is a unitary matrix, hence $\hat{L}\hat{Q}\hat{L}>0$
so that $q_j {\bf e}^{\dagger}_j\hat{L}{\bf
e}_j>0$. Introduce now the "normalized" eigenvectors $\tilde{{\bf
e}}_j={\bf e}_j/\sqrt{\mbox{sgn}{(q_j)}({\bf e}^{\dagger}_j\hat{L}{\bf
e}_j)}$, where sgn$(x)$ stands for the sign function. 
Then it is easy to see that
\[
\tilde{{\bf e}}^{\dagger}_j\hat{L}\tilde{{\bf e}}_j=\mbox{sgn}{(q_j)}
\quad \mbox{and}\quad \tilde{{\bf
e}}^{\dagger}_k\hat{L}\hat{Q}\hat{L} \tilde{{\bf
e}}_j= |q_j|\delta_{kj}
\]

Further introduce the matrix $\hat{T}$ whose columns are vectors
$\tilde{\bf e}_j$ for $j=1,2,...,2n$, and consider
$\hat{T}_L=\hat{T}^{\dagger}\hat{L}$. It is immediately clear that
\[
\hat{T}_L\hat{T}=\mbox{diag}\left(\mbox{sgn}(q_1),...,
\mbox{sgn}(q_{2n})\right)\quad\mbox{and}\quad
\hat{T}_L \hat{Q}\hat{L}\hat{T}=\mbox{diag}\left(|q_1|,...,
|q_{2n}|\right)\equiv\hat{T}_L\hat{T} \mbox{diag}(q_1,...,q_{2n})
\]

The vectors $\tilde{\bf e}_j$ are obviously linearly independent,
 hence the matrices $\hat{T}_L$ and
$\hat{T}$ are nonsingular. This immediately shows that
the matrices $\hat{Q}\hat{L}$ can be diagonalized by a similarity
transformation:
\[
\hat{Q}\hat{L}=\hat{T}\mbox{diag}(q_1,...,q_n)\hat{T}^{-1}
\]
where $\hat{T}$ satisfies:
\[
\hat{T}^{\dagger}\hat{L}\hat{T}=\mbox{diag}\left(\mbox{sgn}(q_1),...,
\mbox{sgn}(q_{2n})\right)
\]

The last matrix is essentially $\hat{L}$ up to a 
permutation of its entries on the main diagonal. This completes the proof.

\section{ Evaluation of the integral Eq.(\ref{coset})}

To evaluate the quoted integral one needs to employ an explicit
parametrisation of the matrices $\hat{T}_0\in
\frac{U(n,n)}{U(n)\times U(n)}$. We follow the paper \cite{VZ} where
it was suggested that the following parametrisation is especially
convenient:
\[
\hat{T}_0=\left(\begin{array}{cc} \sqrt{1+\hat{t}^{\dagger}\hat{t}}&
\hat{t}^{\dagger}\\ \hat{t}& \sqrt{1+\hat{t}\hat{t}^{\dagger}}\end{array}\right)
\]
in terms of complex $n\times n$ matrices
$\hat{t}\,,\,\hat{t}^{\dagger}$. The reason for such a
choice is dictated by an especially simple form of the integration
measure: $d\mu(T_0)=d\hat{t}\,d\hat{t}^{\dagger}$.

Next step is to diagonalise $\hat{t}^{\dagger}$ with help of two
unitary rotations:
$\hat{t}^{\dagger}=u_A^{-1}\hat{\tau}\hat{u}_R\quad\mbox{where}\quad
\hat{u}_{A,R}\in U(n)\quad\mbox{and}\quad
\hat{\tau}=\mbox{diag}(\tau_1,..., \tau_n)$, so that 
$\hat{t}=u_R^{-1}\hat{\tau}^{\dagger}\hat{u}_A$.
It is convenient to write the modulus and the phase of $\tau_k$ 
explicitly: $\tau_k=\sinh{{\psi}_k} e^{i\phi_k}, 
0<\psi<\infty\,,\,0<\phi<2\pi\,,\, k=1,...,n$.

The matrices $\hat{T}_0$  take the form:
\[
\hat{T}_0=\left(\begin{array}{cc}\hat{u}^{-1}_A&0\\0&\hat{u}^{-1}_R
\end{array}\right)
\left(\begin{array}{cc}\cosh{\hat{\psi}} & e^{i\hat{\phi}} \sinh{\hat{\psi}}
\\ e^{-i\hat{\phi}} \sinh{\hat{\psi}} & \cosh{\hat{\psi}}
\end{array}\right)
\left(\begin{array}{cc}\hat{u}_A&0\\0&\hat{u}_R
\end{array}\right)
\]
and $\hat{T}^{-1}_0$ is correspondingly given by:
\[
\hat{T}^{-1}_0=\left(\begin{array}{cc}\hat{u}^{-1}_A&0\\0&\hat{u}^{-1}_R
\end{array}\right)
\left(\begin{array}{cc}\cosh{\hat{\psi}}& -e^{i\hat{\phi}}\sinh{\hat{\psi}}
\\ -e^{-i\hat{\phi}}\sinh{\hat{\psi}} & \cosh{\hat{\psi}}
\end{array}\right)
\left(\begin{array}{cc}\hat{u}_A&0\\0&\hat{u}_R
\end{array}\right) 
\]

One can straightforwardly calculate the integration measure in the new
variables and find:
\begin{eqnarray}\label{measure}
d\hat{t}\,d\hat{t}^{\dagger}&\propto&\prod_{k=1}^n\sinh{2\psi_k}\prod^{n}_{k_1<k_2}
\left(\sinh^2{\psi_{k_1}}-\sinh^2{\psi_{k_2}}\right)^2\prod_{k=1}^n
d\psi_k\,d\phi_k \times d\mu(\hat{u}_A)d\mu(\hat{u}_R) \\
&\propto&\prod^{n}_{k_1<k_2}\left(\lambda_{k_1}-\lambda_{k_2}\right)^2
\prod_{k=1}^n d\lambda_k \, d\phi_k \times d\mu(\hat{u}_A)\,d\mu(\hat{u}_R)
\end{eqnarray}
where $d\mu(\hat{u}_{A,R})$ are normalised 
invariant measures on $U(n)$ and we
introduced: $\lambda_k=\cosh{2\psi_k}\in{[1,\infty)}$ as new variables.

Now we can use the cyclic permutation of the matrices
under the trace sign to rewrite the expression in the exponent of
Eq.(\ref{coset}) in terms of the introduced variables as follows:
 
\begin{eqnarray}
\nonumber && \mbox{Tr}\left[\left(\begin{array}{cc}
\mu_1{\bf 1}_n&\\ & \mu_2^*{\bf 1}_n
\end{array}\right)\hat{T}_0
\left(\begin{array}{cc}\hat{P}_1&\\ & \hat{P}_2
\end{array}\right)\hat{T}^{-1}_0\right]\\ \nonumber
&=&\mbox{Tr}\left(\begin{array}{cc}
\mu_1{\bf 1}_n&\\ & \mu_2^*{\bf 1}_n
\end{array}\right)
\left(\begin{array}{cc}\cosh{\hat{\psi}} & e^{i\hat{\phi}} \sinh{\hat{\psi}}
\\ e^{-i\hat{\phi}} \sinh{\hat{\psi}} & \cosh{\hat{\psi}}
\end{array}\right)
\left(\begin{array}{cc}\hat{P}_A&\\ & \hat{P}_B
\end{array}\right)
\left(\begin{array}{cc}\cosh{\hat{\psi}}& -e^{i\hat{\phi}}\sinh{\hat{\psi}}
\\ -e^{-i\hat{\phi}}\sinh{\hat{\psi}} & \cosh{\hat{\psi}}
\end{array}\right)\\
\nonumber
&=&\mbox{Tr}\left[\hat{P}_A
\left(\mu_1\cosh^2{\hat{\psi}}
-\mu_2^*\sinh^2{\hat{\psi}}\right)\right]
+\mbox{Tr}\left[\hat{P}_B\left(\mu_2^*\cosh^2{\hat{\psi}}-
\mu_1\sinh^2{\hat{\psi}}\right)\right]
\\
&=&\frac{1}{2}(\mu_1+\mu_2^*)\mbox{Tr}\left(\hat{P}_A+\hat{P}_B\right)
+\frac{1}{2}(\mu_1-\mu_2^*)\left(\mbox{Tr}\hat{P}_A\cosh{2\hat{\psi}}
-\mbox{Tr}\hat{P}_B\cosh{2\hat{\psi}}\right)
\end{eqnarray}
where we introduced matrices $\hat{P}_A=\hat{u}_A\hat{P}_1\hat{u}^{-1}_A\,,\,
\hat{P}_B=\hat{u}_B\hat{P}_2\hat{u}^{-1}_B$ having the same
eigenvalues $\hat{q}_1=\mbox{diag}(q_{1,1},...,q_{1,n})$ and 
$\hat{q}_2=\mbox{diag}(q_{2,1},...,q_{2,n})$ as the matrices $\hat{P}_{1,2}$.

We see that the integral of interest is expressed now as:
\begin{eqnarray}
\nonumber
I(\hat{M},\hat{P}_1,\hat{P}_2)&=&\int d\mu(\hat{T})
\exp\left\{i\, N\mbox{Tr}\left(\begin{array}{cc}
\mu_1{\bf 1}_n&\\ & \mu_2^*{\bf 1}_n
\end{array}\right)\hat{T}_0
\left(\begin{array}{cc}\hat{P}_1&\\ & \hat{P}_2
\end{array}\right)\hat{T}^{-1}_0\right\}\\
&\propto&\int_1^{\infty}\prod_k d\lambda_k\, \Delta^2(\hat{\lambda})
e^{\frac{i}{2}N(\mu_1+\mu_2^*)\mbox{Tr}\left(\hat{q}_1+\hat{q}_2\right)}
\\ \nonumber
&\times& \int d\mu(\hat{u}_A)e^{\frac{i}{2}N(\mu_1-\mu_2^*)\mbox{Tr}
\hat{u}_A\hat{P}_1\hat{u}^{-1}_A\hat{\lambda}}
\int d\mu(\hat{u}_R)e^{-\frac{i}{2}N(\mu_1-\mu_2^*)\mbox{Tr}
\hat{u}_R\hat{P}_2\hat{u}^{-1}_R\hat{\lambda}}
\end{eqnarray}
where we used $\hat{\lambda}=\mbox{diag}(\lambda_1,...,\lambda_n)$
and the symbol $\Delta(\hat{\lambda})$ for the corresponding
Vandermonde determinant. 

Two integrals over the (normalized) Haar measure on the unitary group are given by
Harish Chandra-Itzykson-Zuber formula\cite{IZ}:
\[
\int d\mu(\hat{u})e^{\beta\mbox{Tr}
\hat{u}\hat{P}\hat{u}^{-1}\hat{\lambda}}=
\left(\prod_{j=1}^{n-1}j!\right)\beta^{-\frac{n(n-1)}{2}}
\frac{\left.\det{\left[e^{\beta\lambda_kq_l}\right]}
\right|^n_{k,l=1}}{\Delta{(\lambda)}\Delta{(\hat{q})}}
\]
where in our case $\beta=\frac{i}{2}N(\mu_1-\mu_2^*)$ for the first
integral, and for the second one $\beta\to -\beta$.
This gives:
\begin{eqnarray}
\nonumber
I(\hat{M},\hat{P}_1,\hat{P}_2)&\propto&
\left((\mu_1-\mu_2^*)\right)^{-n(n-1)}
\frac{1}{\Delta{(\hat{q}_1)}\Delta{(\hat{q}_2)}}
e^{\frac{i}{2}N(\mu_1+\mu_2^*)\mbox{Tr}\left(\hat{q}_1+\hat{q}_2\right)}
\\ 
&\times&\int_1^{\infty}\prod_k d\lambda_k
\det{\left[e^{\frac{i}{2}N(\mu_1-\mu_2^*)\lambda_kq_{1,l}}\right]^n_{k,l=1}}
\det{\left[e^{-\frac{i}{2}N(\mu_1-\mu_2^*)\lambda_kq_{2,l}}\right]^n_{k,l=1}}
\end{eqnarray}

The last integral can be easily calculated by expanding each 
of the two determinants as:
\[
\det{\left[e^{\pm\beta\lambda_kq_{l}}\right]}^n_{k,l=1}
=\sum_{[S]}(-1)^{[S]}\exp\{\pm\beta\sum_{k=1}^N\lambda_k q_{r_k}\}
\]
where $[S]$ stands for a permutation $r_1,r_2,...,r_n$ of the index
set $1,2,...,n$. Product of two such expansions can be integrated
term by term, and the integrals are convergent due to
$\mbox{Re}\beta>0\,,\, q_{1,i}>0\,,\, q_{2,i}<0$. This gives:
\begin{eqnarray}
\int_1^{\infty}\prod_k d\lambda_k
\det{\left[e^{\beta\lambda_kq_{1,l}}\right]^n_{k,l=1}}
\det{\left[e^{-\beta\lambda_kq_{2,l}}\right]^n_{k,l=1}}
=[-\beta]^{-n}\sum_{[S_1],[S_2]}(-1)^{[S_1]+[S_2]}
\frac{e^{\beta\sum_k q_{1,r_k}-\beta\sum_k q_{2,r_k}}}{\prod_{k=1}^n
\left(q_{1,r_k}-q_{2,l_k}\right)}
\end{eqnarray}
where $[S_1], [S_2]$ are two independent permutations $(r_1,...,r_n)$
and $(l_1,...,l_n)$ of the index set $1,2,...,n$.

Clearly, one can restrict the summation to be taken over the {\it relative}
permutations of the two index sets and multiply the result by $n!$.
The exponential above is invariant with respect to any index
permutation, so it can be taken out of the summation sign and one
recognizes the so-called Cauchy determinant:
\[
\mbox{det}\left(\frac{1}{q_{1,i}-q_{2,j}}\right)_{i,j=1}^n
=
\frac{\Delta({\hat{q}_1})\Delta({\hat{q}_2})}
{\prod_{k_1,k_2}\left(q_{1,k_1}-q_{2,k_2}\right)}
\] 
in the remaining sum. Collecting all the relevant factors together we
 arrive at the final formula:
\[
I(\hat{M},\hat{P}_1,\hat{P}_2)\propto
\left[-i(\mu_1-\mu_2^*)\right]^{-n^2}
\frac{1}{\prod_{k_1,k_2}\left(q_{1,k_1}-q_{2,k_2}\right)}
e^{i N\mbox{Tr}\left(\mu_1\hat{q}_1+
\mu_2^*\hat{q}_2\right)}
\]
up to a constant factor, which can be 
fixed by normalisation in the corresponding equations.

\section{ Negative moments by the Hubbard-Stratonovich transformation}

Let us satisfy ourself that the standard
Hubbard-Stratonovich transformation over the hyperbolic manifold\cite{VZ,VWZ}
produces the same formula Eq.(\ref{2b}). We concentrate
on the simplest nontrivial case $n=1$ for the sake of clarity.
That case was used for a pedagogical introduction into the Hubbard-Stratonovich method 
in the author's lectures in the book \cite{meso} and the notations mainly
follow those lectures.

Our starting point is Eq.(\ref{avmom2}) for $n=1$. 
We introduce the matrices
$\hat{A}=\hat{L}^{1/2}\hat{Q}\hat{L}^{1/2}$ so that
$\mbox{Tr}\hat{A}^2=\mbox{Tr}\hat{Q}\hat{L}\hat{Q}\hat{L}$ and 
$\mbox{Tr}{A}={\bf S}^{\dagger}_1{\bf S}_1-{\bf S}^{\dagger}_2{\bf
S}_2\,\,,\,\,\mbox{Tr}{AL}={\bf S}^{\dagger}_1{\bf S}_1+{\bf S}^{\dagger}_2{\bf
S}_2$. Remembering $\mu_{1,2}=\mu\pm(\omega/2+i\delta)$ we can
express all terms appearing in the exponent of Eq.(\ref{avmom2})
in terms of $\hat{A}$:
\[
\frac{i}{2}\left(\mu_1{\bf S}^{\dagger}_1{\bf S}_1-
\mu_2{\bf S}^{\dagger}_2{\bf
S}_2\right)=\frac{i}{2}\mu\mbox{Tr}\hat{A}+
\frac{1}{2}(\delta-i\omega/2)\mbox{Tr}\hat{A}\hat{L}
\]
The Hubbard-Stratonovich transformation is the identity:
\begin{eqnarray}\label{Hub}
&&\exp\left\{-\frac{1}{8N}\mbox{Tr}\hat{A}^2-\frac{1}{2}(\delta-i\omega/2)
\mbox{Tr}\hat{A}\hat{L}\right\}  \\
\nonumber &\propto& e^{N(\delta-i\omega/2)^2}\int d\hat{Q}
\exp\left\{-\frac{N}{2}\mbox{Tr}\hat{Q}^2-
iN(\delta-i\omega/2) \mbox{Tr}(\hat{Q}\hat{L})-
\frac{i}{2}\mbox{Tr}\hat{Q}\hat{A}\right\}
\end{eqnarray}

Despite looking as an innocent gaussian integral the identity is
very nontrivial, since the convergency arguments force one to choose
the following "hyperbolic manifold" of 
the matrices $\hat{Q}$ as the integration
domain: 
\begin{eqnarray}\label{hypman}
\hat{Q}&=&\hat{T}^{-1}\left(\begin{array}{cc}p_1&0\\0&p_2\end{array}\right)
\hat{T}\quad,\quad 
\hat{T}=\left(\begin{array}{cc}\cosh{\theta} & e^{i\phi}\sinh{\theta}\\
e^{-i\phi}\sinh{\theta}& \cosh{\theta}\end{array}\right),\\
\nonumber d\hat{Q}&\propto& (p_1-p_2)^2\sinh{2\theta} \, dp_1
dp_2\,d\theta \, d\phi, \\ \nonumber && -\infty<\mbox{Re}(p_{1,2})<\infty\,,\,
\mbox{Im}(-p_1,p_2)>0\,\,,\,0\le
\theta<\infty\,,\, 0<\phi<2\pi
\end{eqnarray}

A detailed discussion of the convergency problems and of the
above identity Eq.(\ref{Hub}) can be found, e.g., in the book
\cite{Haake} and in the mentioned lectures\cite{meso}. 

Substituting such an identity back to Eq.(\ref{avmom2}) and changing
the order of integrations over $\hat{Q}$ and ${\bf S}_{1,2}$ we
see that it can be processed as follows:
\begin{eqnarray}
&&e^{N(\delta-i\omega/2)^2}\int d\hat{Q}
\exp\left\{-\frac{N}{2}\mbox{Tr}\hat{Q}^2-
iN(\delta-i\omega/2) \mbox{Tr}(\hat{Q}\hat{L})\right\}\\ 
\nonumber
&&\times \int d{\bf S}d{\bf S}^{\dagger}
\exp\left\{\frac{i}{2}\mu {\bf S}^{\dagger}\hat{L}{\bf S}-
\frac{i}{2} {\bf S}^{\dagger}\hat{L}^{1/2}
\hat{Q}^T\hat{L}^{1/2}{\bf S}\right\} \\
\nonumber 
&&\propto e^{N(\delta-i\omega/2)^2}
\int d\hat{Q}\left[\det{(\mu\hat{I}_n-\hat{Q})}\right]^{-N}\,\,
\exp\left\{-\frac{N}{2}\mbox{Tr}\hat{Q}^2-
iN(\delta-i\omega/2) \mbox{Tr}(\hat{Q}\hat{L})\right\}\\
\nonumber
&\propto& e^{N(\delta-i\omega/2)^2}
\int dp_1 \int dp_2 \frac{(p_1-p_2)^2}{\left[(\mu-p_1)(\mu-p_2)\right]^N}
\exp\left\{-\frac{N}{2}(p_1^2+p_2^2)\right\}\\
\nonumber &\times&\int _0^{2\pi}d\phi\int_0^{\infty}d\theta \sinh{2\theta}
\exp\left\{-iN(\delta-i\omega/2)\mbox{Tr}\left[
\left(\begin{array}{cc}p_1&0\\0&p_2\end{array}\right)
\hat{T}\hat{L}\hat{T}^{-1}\right]\right\}
\end{eqnarray}
where we introduced the notation ${\bf S}=({\bf S}_1,{\bf S_2})$
and used $\mbox{Tr}\hat{Q}\hat{A}={\bf S}^{\dagger}\hat{L}^{1/2}\hat{Q}^T\hat{L}^{1/2}{\bf S}$.

Using the explicit parametrisation for $\hat{T}$ it is easy to verify
that: 
\[
\mbox{Tr}\left[
\left(\begin{array}{cc}p_1&0\\0&p_2\end{array}\right)
\hat{T}\hat{L}\hat{T}^{-1}\right]=(p_1-p_2)\cosh{2\theta}
\]
which allows one to perform the integration over $\theta$. The expression
above is therefore reduced to:
\begin{equation}\label{l}
\frac{e^{N(\delta-i\omega/2)^2}}{\delta-i\omega/2}
\int dp_1 \int dp_2 \frac{p_1-p_2}{\left[(\mu-p_1)(\mu-p_2)\right]^N}
\exp\left\{-\frac{N}{2}(p_1^2+p_2^2)-iN(\delta-i\omega/2)(p_1-p_2)
\right\}
\end{equation} 

Remembering $\mbox{Im}{p}_1<0\,,\,\mbox{Im}{p}_2>0$ we can use the
identity:  
\[
\left[(\mu-p_1)(\mu-p_2)\right]^{-N}\propto 
\int_0^{\infty}dq_1\int_0^{\infty}dq_2 (q_1q_2)^{N-1}
\exp\{i\left[(\mu-p_1) q_1-(\mu-p_2) q_2\right]\}
\]
and further rewrite the integral Eq.(\ref{l}) as:
\begin{eqnarray}
&\propto&\frac{e^{N(\delta-i\omega/2)^2}}{\delta-i\omega/2} 
\int_0^{\infty}dq_1 \int_0^{\infty}dq_2 (q_1q_2)^{N-1}
e^{i\mu(q_1-q_2)}
\\ \nonumber &\times&
\int dp_1 \int dp_2 (p_1-p_2)
\exp\left\{-\frac{N}{2}(p_1^2+p_2^2)-iN(\delta-i\omega/2)(p_1-p_2)-
i(p_1q_1-p_2q_2)\right\}
\end{eqnarray}
 
The next step is to use the chain of identities:
\begin{eqnarray}
&&
\int dp_1 \int dp_2 (p_1-p_2)
\exp\left\{-\frac{N}{2}(p_1^2+p_2^2)-iN(\delta-i\omega/2)(p_1-p_2)
-i(p_1q_1-p_2q_2)\right\}
\\ \nonumber &\propto&
\left(\frac{\partial}{\partial q_1}+
\frac{\partial}{\partial q_2}\right) \int dp_1  \int dp_2
\exp\left\{-\frac{N}{2}(p_1^2+p_2^2)-ip_1
\left[N(\delta-i\omega/2)+q_1\right]+ip_2
\left[N(\delta-i\omega/2)+q_2\right] \right\}\\
\nonumber &\propto& 
\left(\frac{\partial}{\partial q_1}+\frac{\partial}{\partial q_2}\right)
\exp\left\{-\frac{1}{2N}(q_1^2+q_2^2)
-(\delta-i\omega/2)
(q_1+q_2)-N(\delta-i\omega/2)^2\right\}  
\end{eqnarray}
and observe that 
\[
e^{i\mu(q_1-q_2)}
\left(\frac{\partial}{\partial q_1}+\frac{\partial}{\partial q_2}\right)
F(q_1,q_2)\propto \left(\frac{\partial}{\partial q_1}+
\frac{\partial}{\partial q_2}\right)
\left[e^{i\mu(q_1-q_2)}F(q_1,q_2)\right]
\]
The last formula allows one to integrate by parts
over $q_1,q_2$ and in this way to get rid of the derivatives. 
The boundary terms vanish for $N>1$, the application of the operator
$\left(\frac{\partial}{\partial q_1}+\frac{\partial}{\partial q_2}\right)$
to $(q_1q_2)^{N-1}$ produces the term $(q_1q_2)^{N-2}(q_1+q_2)$
 and the resulting expression concides with that given in eq.(\ref{2b}) for $n=1$.

For $n>1$ the equivalence can be shown along essentially the same lines, but
calculations become cumbersome and require the use of the procedure similar to that
outlined in the Appendix C.

\end{document}